\documentclass{ws-jai}

\usepackage[separate-uncertainty=true, retain-explicit-plus=true]{siunitx}
\DeclareSIUnit\lbs{lbs}
\DeclareSIUnit\inch{in}

\usepackage[flushleft]{threeparttable}

\begin{document}

\catchline{}{}{}{}{} 

\markboth{F.~Kislat et al.}{The X-Calibur truss}

\title{Design of the telescope truss and gondola for the balloon-borne X-ray polarimeter X-Calibur}

\author{Fabian Kislat$^{\dagger*}$, Banafsheh Beheshtipour$^\dagger$, Paul Dowkontt$^\dagger$, Victor Guarino$^{\dagger|}$, R. James Lanzi$^\ddagger$, Takashi Okajima$^\S$, Dana Braun${\dagger}$, Scott Cannon$^{\&}$, Gialuigi De Geronimo$^\P$, Scott Heatwole$^\ddagger$, Janie Hoorman$^\dagger$, Shaorui Li$^\P$, Hideyuki Mori$^\S$, Christopher M. Shreves$^\ddagger$, David Stuchlik$^\ddagger$ and Henric Krawczynski$^\dagger$}

\address{
$^\dagger$Department of Physics and McDonnell Center for the Space Sciences, Washington University in St. Louis, Saint Louis, MO 63130, USA\\
$^|$Guarino Engineering Services, 1134 S. Scoville Ave, Oak Park, IL 60304, USA\\
$^\ddagger$NASA Wallops Flight Facility, Wallops Island, VA 23337, USA\\
$^\S$NASA Goddard Space Flight Center, Greenbelt, MD 20771, USA\\
$^{\&}$Physical Science Laboratory, New Mexico State University, Las Cruces, NM 88003, USA\\
$^\P$Brookhaven National Laboratory, Upton, NY 11973, USA
}

\maketitle

\corres{$^*$Corresponding author. Email: fkislat@physics.wustl.edu}

\begin{history}
\received{(to be inserted by publisher)};
\revised{(to be inserted by publisher)};
\accepted{(to be inserted by publisher)};
\end{history}

\begin{abstract}
X-ray polarimetry has seen a growing interest in recent years.
Improvements in detector technology and focusing X-ray optics now enable sensitive astrophysical X-ray polarization measurements.
These measurements will provide new insights into the processes at work in accreting black holes, the emission of X-rays from neutron stars and magnetars, and the structure of AGN jets.
X-Calibur is a balloon-borne hard X-ray scattering polarimeter.
An X-ray mirror with a focal length of \SI{8}{\m} focuses X-rays onto the detector, which consists of a plastic scattering element surrounded by Cadmium-Zinc-Telluride detectors, which absorb and record the scattered X-rays.
Since X-rays preferentially scatter perpendicular to their polarization direction, the polarization properties of an X-ray beam can be inferred from the azimuthal distribution of scattered X-rays.
A close alignment of the X-ray focal spot with the center of the detector is required in order to reduce systematic uncertainties and to maintain a high photon detection efficiency.
This places stringent requirements on the mechanical and thermal stability of the telescope structure.
During the flight on a stratospheric balloon, X-Calibur makes use of the Wallops Arc-Second Pointer (WASP) to point the telescope at astrophysical sources.
In this paper, we describe the design, construction, and test of the telescope structure, as well as its performance during a 25-hour flight from Ft.\ Sumner, New Mexico.
The carbon fiber-aluminum composite structure met the requirements set by X-Calibur and its design can easily be adapted for other types of experiments, such as X-ray imaging or spectroscopic telescopes.
\end{abstract}


\section{Introduction}
X-ray astronomy has led to spectacular insights about the most extreme objects in our Universe, such as, amongst others, black holes and neutron stars.
While polarization is measured routinely at longer wavelengths, for example with optical and radio telescopes, only few X-ray polarization measurements have been carried out so far.
Most X-ray observations to date have been restricted to imaging, spectroscopic and timing measurements.
X-ray polarimetry is expected to give a new impulse to the field of X-ray astronomy, by providing two new observables: the linear polarization fraction and position angle of the radiation expressed by the Stokes parameters $Q$ and $U$ which give the intensity of linear polarized light along two directions rotated by \SI{45}{\degree} relative to each other.
By measuring the energy-dependent polarization properties of X-rays, one obtains information about the location of the X-ray emission, their propagation through the curved spacetime and/or magnetized plasma, scattering locations, and polarization dependent absorption processes (see \citet{1988ApJ...324.1056M}, \citet{1997SSRv...82..309L}, \citet{2009ASSL..357..589W}, \citet{2010xpnw.book.....B}, and \citet{2011APh....34..550K} for reviews).
In other words, X-ray polarization carries geometric information about compact sources such as black holes and neutron stars, which are only a few femto degrees across in the sky, and hence cannot be resolved at any wavelength in the foreseeable future.
This additional information will allow us to test many of the models of X-ray sources that have been developed since the discovery of extra-solar X-ray emission from Scorpius X-1 by \citet{1962PhRvL...9..439G}.

While this potential of X-ray polarimetry has long been realized, so far only one dedicated X-ray polarimeter has been flown on a satellite.
The Bragg polarimeter on board the OSO-8 satellite launched in 1978 measured a polarization fraction of the Crab Nebula of about~$20\%$ at energies of~\SI{2.6}{\keV} and~\SI{5.2}{\keV}~\citep{1978ApJ...220L.117W}.
Since then, only a few more X-ray polarization measurements have been published.
In 2008, the instruments SPI and IBIS on board the INTEGRAL satellite reported polarization fractions of the Crab Nebula of~$46\pm10\%$~\citep{2008Sci...321.1183D} and~${>}72\%$~\citep{2008ApJ...688L..29F}, respectively, with the polarization direction aligned with the X-ray jet.
The balloon-borne X-ray polarimeter PoGOLite reported a polarization fraction of the phase-averaged Crab emission of $21 \pm 10\%$ in the energy range from 20 to \SI{240}{keV}~\citep{2016MNRAS.456L..84C}.
For the stellar mass black hole Cygnus X-1, a polarization fraction of~$40\pm10\%$ in the $230$ to \SI{400}{\keV} range and ${>}75\%$~\citep{2012ApJ...761...27J} and~$67\pm30\%$~\citep{2011Sci...332..438L} at higher energies have been reported.
For a number of Gamma-Ray Bursts, tentative evidence for polarized emission has been published~\citep{2003Natur.423..415C,2007ApJS..169...75K,2011ApJ...743L..30Y}, but the measurements are plagued by large statistical and systematic uncertainties.
More recently, NASA has selected two photo-electric effect polarimeters for Phase A studies: IXPE~\citep{2013SPIE.8859E..08W} and PRAXyS~\citep{2014SPIE.9144E..0NJ}.
In Europe, the imaging photo-electric effect polarimeter XIPE is being studied as a candidate for a medium-size mission~\citep{2013ExA....36..523S}.

\begin{figure}
  \centering
  \includegraphics[width=.6\textwidth]{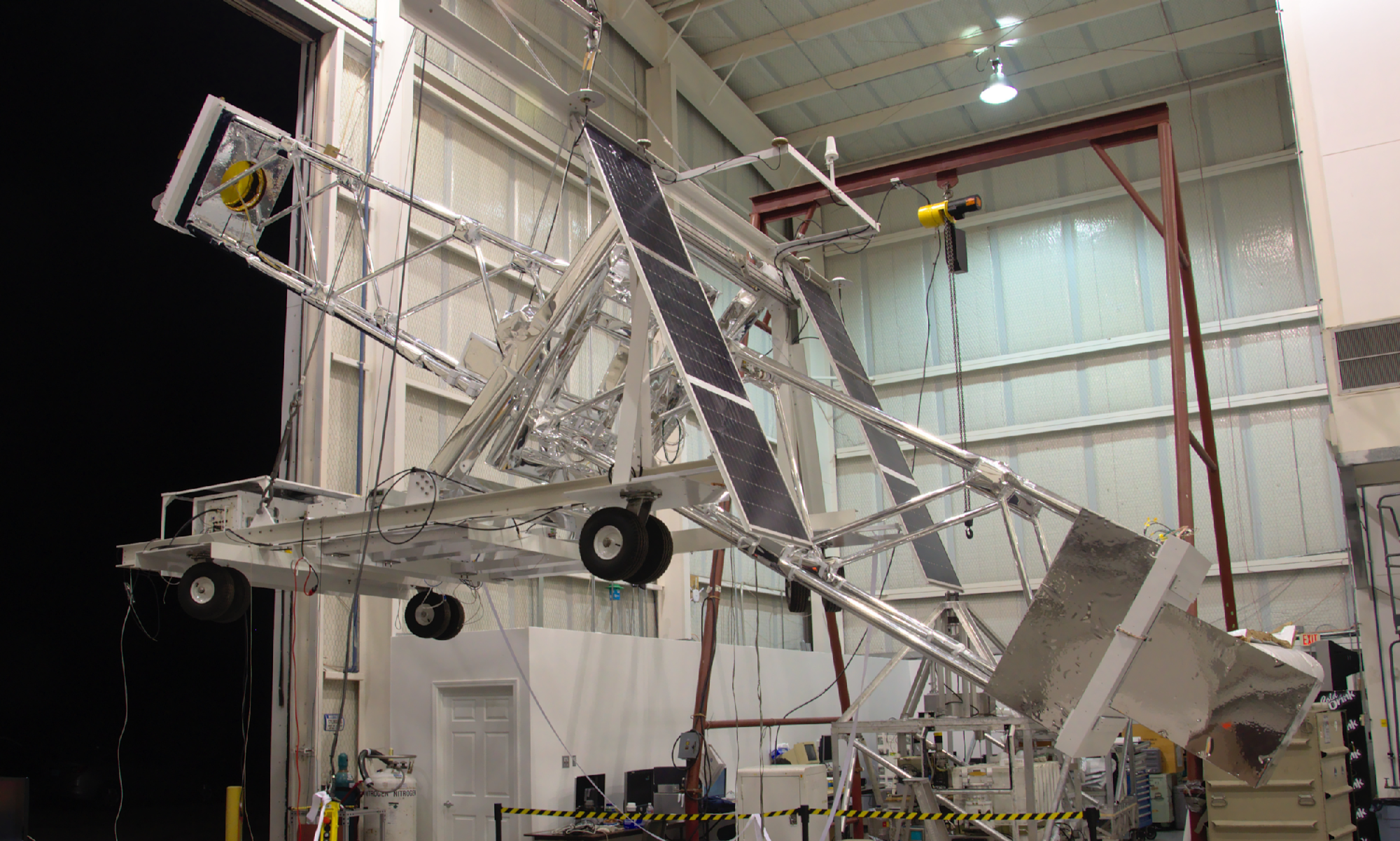}
  \caption{The X-Calibur/InFOCuS/WASP telescope during a night-time pointing test at Ft.\ Sumner (NM) in preparation of the 1-day balloon flight in September, 2016. The detector assembly is inside the enclosure on the right-hand side of the truss structure. The X-ray mirror is on the left side of the truss.}
  \label{fig:telescope}
\end{figure}

X-Calibur~\citep{2014JAI.....340008B} is a balloon-borne scattering polarimeter for the energy range from $20$ to \SI{60}{\keV}.
It was flown on a one-day stratospheric balloon flight from Ft.\ Sumner (NM) in 2014 in the focal plane of the InFOCuS telescope.
During the flight, valuable background data were acquired, but due to a failure of the pointing system, no cosmic X-ray sources were observed.
We have upgraded X-Calibur and successfully flown it in fall 2016 using the Wallops Arc-Second Pointer system (WASP) and a new telescope structure, as shown in Fig.~\ref{fig:telescope}.
Core requirements when designing this new structure were its mechanical stiffness and limits on thermal deformations derived from the provision that the focal spot of the X-ray mirror should remain within \SI{3}{\mm} of the center of the detector assembly.
In this paper, we describe the design and construction of the new telescope truss, the systems used to monitor alignment of the X-ray mirror with the detector during flight, and the flight performance of the system.
The telescope design can easily be adapted for other uses, after adjusting for focal length and balance.

The paper is structured as follows: In Section~\ref{sec:xcalibur} we briefly describe the X-Calibur polarimeter.
Section~\ref{sec:wasp} gives an overview over the Wallops Arc-Second Pointer (WASP).
In Section~\ref{sec:requirements} we describe the design requirements of the truss structure, including the pointing budget and thermal and mechanical considerations.
In Section~\ref{sec:design} we describe the mechanical design and the systems we developed to monitor mirror-to-detector alignment during the flight.
The thermal and mechanical analysis is described in Section~\ref{sec:analysis}.
In Section~\ref{sec:tests} we give an account of the component tests performed during the design phase and the pre-flight tests to qualify the assembled structure for the balloon flight.
In Section~\ref{sec:flight} we give an account of the in-flight performance during the September 2016 flight, and finally Section~\ref{sec:summary} summarizes the paper.

\section{The X-ray polarimeter X-Calibur}\label{sec:xcalibur}
\begin{figure*}
  \centering
  \includegraphics[width=.8\textwidth]{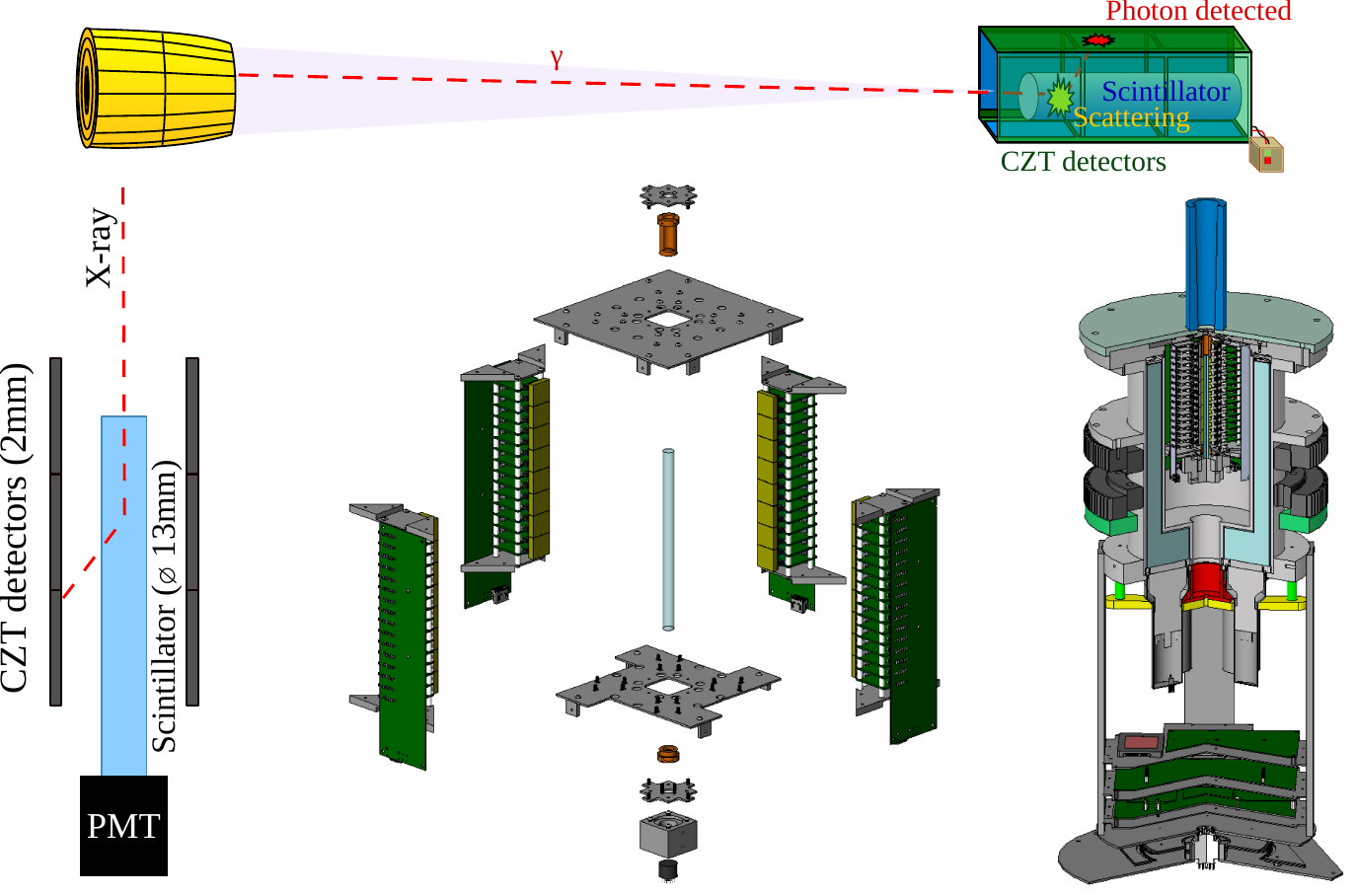}
  \caption{The detector principle of X-Calibur. \emph{Top:} The InFOCuS X-ray mirror (left) focuses X-rays onto a plastic scintillator rod, in which they scatter. The scattered X-rays are then detected by CdZnTe detectors surrounding the scattering rod. \emph{Left:} The \SI{4.3}{\inch} long, \SI{13}{\mm} diameter scintillator rod is surrounded by \SI{2}{\mm} thick, $2\times\SI{2}{\cm\squared}$ detectors, three on each side. A PMT at the end of the scintillator detects the light generated in the rod when a photon scatters. This can be used as a coincidence signal. \emph{Center:} Exploded view of the detector configuration as flown during the first flight with eight detectors on each side. \emph{Right:} The detector is embedded inside an active CsI(Na) shield. The entire assembly is rotated during flight in order to reduce systematic uncertainties.}
  \label{fig:xcalibur}
\end{figure*}

X-Calibur is a hard X-ray scattering polarimeter.
Its design, laboratory testing, and results from the first flight are described in~\citet{2013APh....41...63G}, and \citet{2014JAI.....340008B, beilicke_xcalibur_2015}.
Here, we summarize the design and highlight changes that have been made since the first flight.
The InFOCuS grazing incidence X-ray mirror focuses X-rays onto a \SI{13}{\mm}-diameter plastic scintillator rod, in which the X-rays scatter.
The scattered X-rays are detected by an array of Cadmium-Zinc-Telluride (CZT) detectors surrounding the scattering rod, as shown in Fig.~\ref{fig:xcalibur}.
Linearly polarized photons scatter preferentially perpendicular to the polarization direction.
One can then assign Stokes parameters $I$, $Q$, and $U$ to each detected photon, and determine the polarization of the beam from the sum of the Stokes parameters of all events~\citep{2015APh....68...45K}.

\begin{figure}
 \centering
 \includegraphics[width=.3\textwidth]{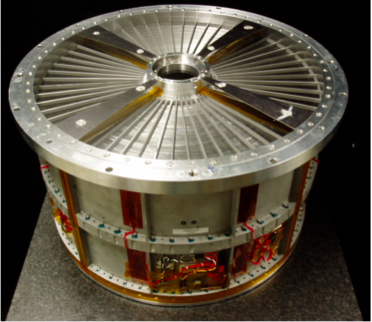}
 \caption{The InFOCuS X-ray mirror.}
 \label{fig:mirror}
\end{figure}

The Wolter-type grazing incidence mirror of the InFOCuS telescope~\citep[see Fig.~\ref{fig:mirror}]{2005SPIE.5900..217O} consists 255 nested Aluminum shells with Tungsten/Carbon multi-layer coatings to enable an energy bandpass up to \SI{60}{\keV}.
It has a focal length of \SI{8}{\m} and an effective area of $93/60/\SI{30}{\square\cm}$ at $20/30/\SI{40}{\keV}$, respectively.
The field of view is \SI{8}{\arcmin} at \SI{20}{\keV}, and the \SI{75}{\percent} containment radius of the focal spot is \SI{3.5}{\arcmin}, corresponding to a width of about \SI{8}{\mm}.

The focal spot of the mirror is centered at the tip of a \SI{4.3}{\inch} long, \SI{13}{\mm}-diameter EJ-200 plastic scintillator.
The scintillator is surrounded by twelve $2\times\SI{2}{\cm\squared}$ CZT detectors with a thickness of~\SI{2}{\mm}.
The number of detectors was reduced compared to the first flight keeping only the detectors with the best signal-to-background ratio.
Each detector is subdivided into 64 pixels with a pitch of \SI{2.5}{\mm}, which are read by two 32-channel NCI ASICs developed at Brookhaven National Laboratory~\citep{2007NIMPA.579..371W}.
Light signals in the scintillator rod are detected by a Hamamatsu R7600U-200 PMT with a high quantum efficiency super-bialkali photo cathode.
These signals can be used in the data analysis to suppress background by about one order of magnitude.

As described above, the polarization fraction and angle are inferred from the azimuthal distribution of an ensemble of photons detected in the CZT detectors surrounding the scattering stick.
Systematic uncertainties may, thus, arise from differences in the detection efficiencies and thresholds of the different CZT detectors.
In order to reduce these effects, the entire assembly is continuously rotated around the optical axis.

An \SI{8}{\m}-long telescope structure holds the mirror and the polarimeter on either end.
The truss structure consists of two halves that are connected to a set of gimbal frames which can be turned in pitch and yaw by the Wallops Arc-Second Pointer (WASP) system, in order to point the telescope at astrophysical sources during the flight.

\section{The Wallops Arc-Second Pointer}\label{sec:wasp}
X-Calibur utilizes the Wallops Arc-Second Pointer (WASP), a high-accuracy attitude control system for balloon-borne payloads developed at NASA's Wallops Flight Facility~\citep{stuchlik_d_2015}.
A key design aspect of WASP was the flexibility to accommodate a variety of science instruments, and the reusability of major components.
The WASP system points an instrument using a gondola mounted pitch/yaw articulated gimbal.
The range of motion of the yaw-gimbal is purposely minimized to reduce kinematic coupling during fine pointing.
Thus, the gondola itself is suspended beneath a standard NASA Rotator to provide large angle azimuth targeting and coarse azimuth stabilization.

\begin{figure}
  \centering
  \includegraphics[width=0.4\textwidth]{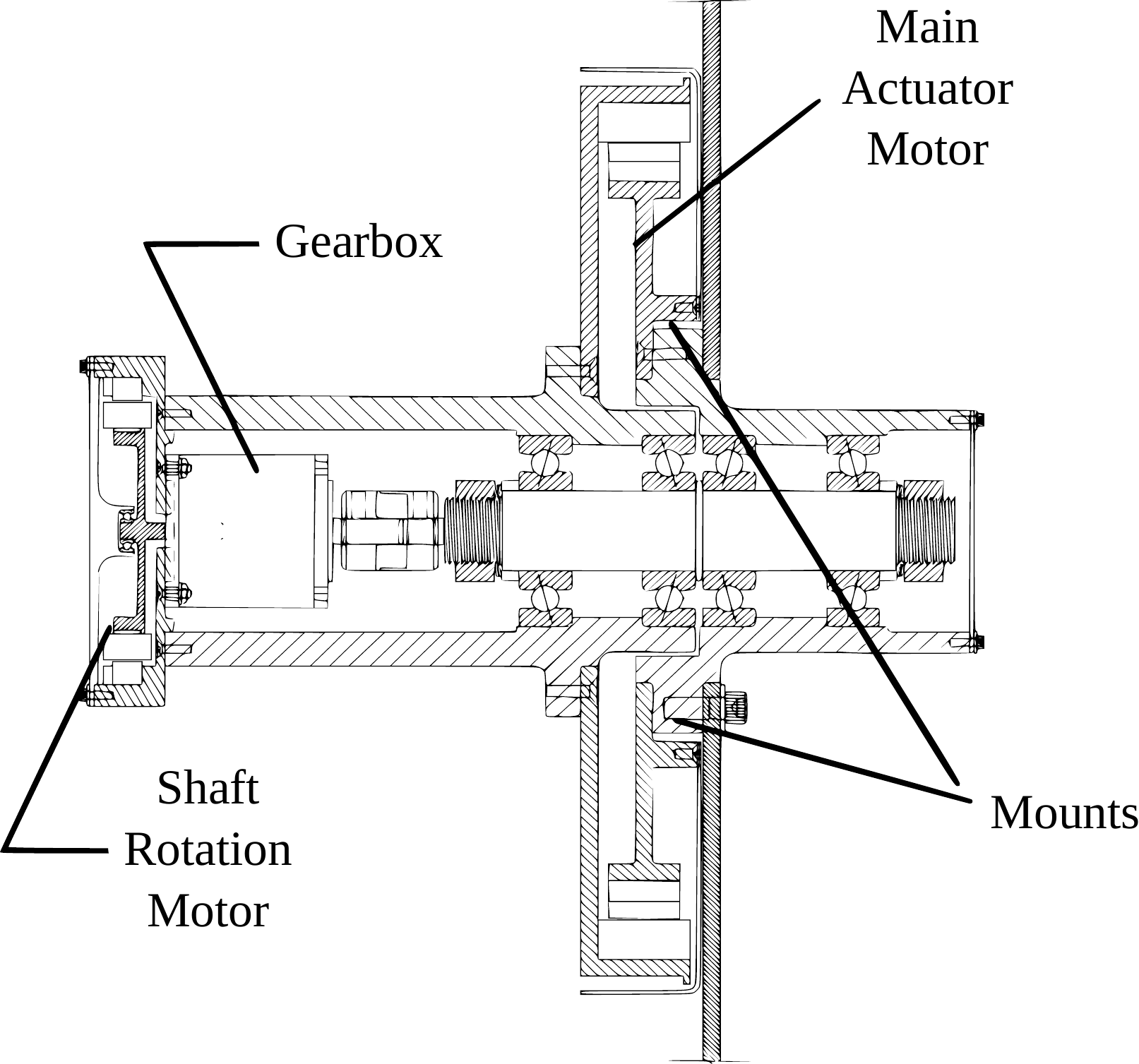}
  \caption{WASP actuator motor detail. The central shaft is itself rotated to eliminate static friction, and the shafts of opposite sides of the gimbal are counter-rotated in order to reduce residual kinetic friction.}
  \label{fig:wasp-hub-motor}
\end{figure}

Sub-arc-second pointing is enabled by the mechanical design of the WASP gimbal hubs (see Fig.~\ref{fig:wasp-hub-motor}).
A pair of hubs on opposing sides of the gimbal is used to establish each articulated axis of rotation.
Each hub uses high-precision angular contact bearings to float the rotor side and stator side of the hub on a central shaft.
The central shaft in each hub is itself rotated by a small diameter torque motor through a gear-box to eliminate static friction.
The shafts in each hub pair are counter-rotated in an attempt to reduce the residual kinetic friction that must be corrected by the control system.

A large-diameter brushless DC torque motor is used to provide the torque for each control axis.
The current in the three electrical phases in each of the control motors is commutated in software and set using power Op-Amps in a motor interface circuit.
To accomplish this, rotor-to-stator angle for each torque motor is determined through the use of resolvers, one for each control axis. 

Instrument attitude is computed by integrating incremental angles output from a LN251 system, whose use in WASP is limited to that of a relatively low-cost, high-quality inertial-rate-unit.
Control torques are computed using a modified Proportional-Integral-Derivative control law in each axis. 

The WASP system also includes a star tracker camera, the Camera Attitude Reference Determination System (CARDS), which has been developed specifically for balloon flight applications.
Its quaternion output is blended with the integrated attitude solution from the LN251 utilizing a 6-state Extended Kalman Filter (EKF).
The WASP EKF uses the output from the CARDS to provide direct updates to the integrated attitude solution and to provide running estimates of rate biases inherent in the output of the LN251.
In addition to accepting quaternions, the WASP EKF is capable of blending science-target unit-vector data into the attitude solution to provide autonomous corrections for offsets between the CARDS and the science instrument (``software shims'').

\section{Design requirements}\label{sec:requirements}
\subsection{Pointing and alignment budget}\label{sub:requirements:pointing}
The design of the telescope truss for X-Calibur was driven by the scientific requirements of the mission.
In order to maintain systematic uncertainties at an acceptable level and maintain high photon detection efficiencies, the focal spot must not be offset from the center of the scattering rod by more than~\SI{3}{\mm}, corresponding to a misalignment angle of \SI{77}{\arcsecond}.
This places stringent requirements on the alignment and mechanical as well as thermal stability of the system.

\begin{figure}
  \centering
  \includegraphics[width=.5\textwidth]{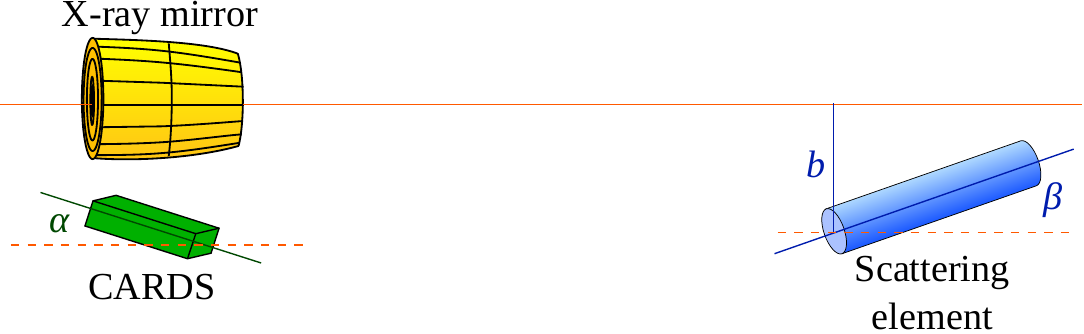}
  \caption{Illustration of the alignment budget for X-Calibur. Three factors are affecting alignment: The misalignment between the CARDS star tracker and the X-ray mirror ($\alpha$); the misalignment between the optical axis of the mirror and the scattering element ($\beta$); and an offset between the optical axis and the scattering element ($b$).}
  \label{fig:alignmentbudget}
\end{figure}

Possible sources of misalignment are illustrated in Fig.~\ref{fig:alignmentbudget}.
Three variables have to be taken into consideration:
\begin{itemlist}
 \item the angle $\alpha$ between the optical axis of the X-ray mirror and the star tracker;
 \item the angle $\beta$ between the scattering element and the optical axis of the mirror;
 \item and the offset $b$ between the optical axis of the X-ray mirror and the center of the scattering element.
\end{itemlist}
Keeping the offset $b$ within acceptable limits throughout the flight dictated the requirements on the mechanical stiffness and thermal stability of the truss.
Additionally, it means that the scattering slab must be on the polarimeter rotation axis, and that this axis must be aligned with the optical axis of the X-ray mirror.

Pre-flight alignment was achieved using a multi-step procedure.
Two cameras can be mounted to the central bore-hole of the X-ray mirror: a forward-looking star camera to observe the night sky, and a backward-looking camera focused at the detector assembly.
The mounting of both cameras was designed such that their orientation with respect to the mirror optical axis can be reproduced with a precision of about \SI{15}{\arcsecond}.

\begin{figure}
  \centering
  \includegraphics[width=.6\textwidth]{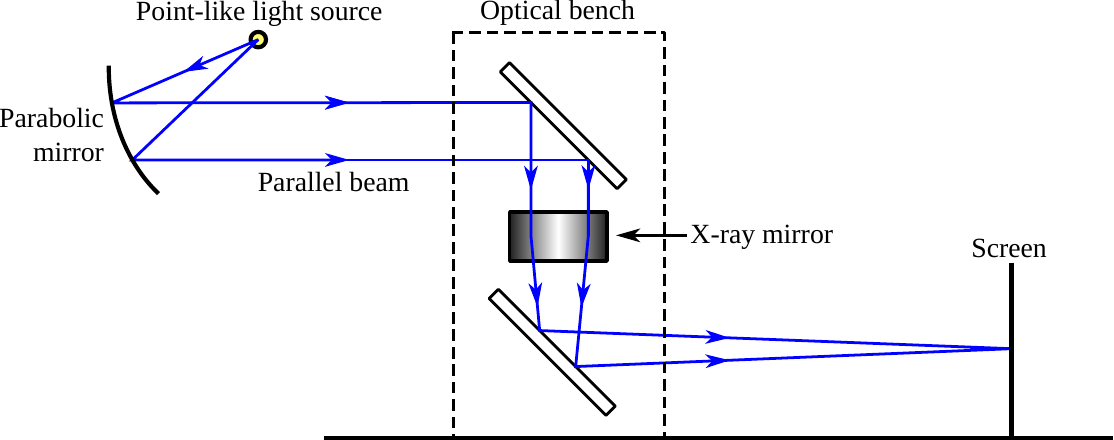}
  \caption{Setup to calibrate the forward-looking camera used to measure the alignment between X-ray mirror and star tracker, as well as the backward-looking camera used to align the mirror with the detector.}
  \label{fig:mirror-alignment}
\end{figure}

Both cameras were calibrated using the setup shown in Fig.~\ref{fig:mirror-alignment}.
The X-ray mirror was illuminated with a parallel beam of light created by reflecting a point-like light source off of a parabolic mirror.
The orientation of the mirror is then adjusted using micrometer screws to align its optical axis with the light beam.
The X-ray mirror focuses the optical light beam in the same way as a beam of X-rays.
The backward-looking camera is mounted to the mirror and the image of the light source on a screen created by the X-ray mirror is observed.
When aligning the mirror with the detector, as described below, the mirror is shimmed such that the detector center will be at the location of the image of the optical beam.
Similarly, the forward-looking camera is mounted to the mirror, and the incoming parallel beam of light is observed.
When cross-calibrating the star tracker and the mirror, as described below, the pointing system will be adjusted to point the telescope such that a target star will appear at the same location as the parallel beam.

After installation of the mirror in the telescope, it was shimmed mechanically in order to position its focal spot at the center of the detector assembly in order to cancel the offset $b$.
The backward-looking camera calibrated in the previous step was used to determine the focal spot location.
By comparing images with and without the scattering element installed in the polarimeter it was verified, that the angle $\beta$ between the optical axis and the scattering element was less then \SI{0.5}{\degree}, resulting in an offset of the rear end of the scintillator of less than \SI{1}{\mm} from the front.
Additionally, the polarimeter was rotated during this procedure to verify that the excentricity of the assembly was less than \SI{1}{\mm}.

In the second step, a flat mirror mounted perpendicular to the optical axis of the mirror and a theodolite were used to mechanically align the X-ray mirror and the CARDS star tracker.
After this coarse alignment, the forward-looking star camera was mounted to the X-ray mirror.
During a night-time test, images were taken with both this forward-looking camera and CARDS.
From these images a correction quaternion was calculated as an offset in the pointing algorithm, resulting in an absolute pointing precision of the mirror of \SI{15}{\arcsecond}.
The mechanical misalignment between mirror and star tracker was \SI{13}{\arcmin}.
It was verified that by applying these software shims, a star was centered in the pixel of the forward-looking camera that corresponds to the mirror optical axis.
This star was tracked for an extended period and the procedure was repeated with a second target for verification.
During the test, up to 9 stars were tracked by CARDS.
Improvements to the cross-calibration procedure are currently being developed in order to achieve an absolute pointing accuracy better than \SI{1}{\arcsecond}.

\subsection{Mechanical and thermal considerations}
The pre-flight alignment has to be maintained throughout the flight, resulting in the following requirements on the truss design:
\begin{itemlist}
 \item While tracking a source, the telescope truss elevation changes from about \SI{30}{\degree} to \SI{65}{\degree} resulting in changing gravitational forces, possibly resulting in a bending of the structure with respect to its horizontal configuration. During the design phase, a finite-element model of the truss was used to study deformations of the truss as its angle with respect to the gravity vector changes. Deflections up to \SI{3}{\mm} can be tolerated, but the design goal was to achieve a stability of \SI{1}{\mm} or better.
 \item At float altitude, the truss will be significantly colder than on the ground, and in particular in Ft.\ Sumner the temperatures will differ significantly between day and night. Furthermore, temperature gradients along the truss are significantly larger than on the ground. During the engineering phase, detailed thermal simulations of different extreme cases were carried out, predicting the temperature at various positions on the truss. These temperature predictions were then fed into the finite element model of the truss in order to calculated resulting deformations.
 \item A pointing knowledge better than \SI{1}{\mm} is required in the data analysis. This lead to the design of two redundant alignment monitoring systems described in Section~\ref{sub:design:monitoring}.
\end{itemlist}

The maximum width and height of the structure are dictated by the dimensions of the size of the WASP gimbal frames, as well as the requirement that the truss must be able to elevate from horizontal to an elevation of \SI{65}{\degree}, as well as slew $\pm\SI{7}{\degree}$ in yaw at all elevations.
The slewing requirements also had to be taken into account when designing the gondola.
Furthermore, the pointing system places requirements on the rigidity of the pointed structure.
In order to prevent the control system from coupling with the structure, the first resonant mode of the truss must have a frequency greater than~\SI{10}{\Hz}.

Besides these constraints due to scientific requirements, the truss was also designed to be as lightweight and balanced as possible.
Weight limitations arise from the fact that (a) WASP can point telescopes weighing up to \SI{1500}{\lbs}, and (b) in order to perform sensitive X-ray observations of astrophysical sources, X-Calibur has to fly as high as possible in order to minimize absorption of X-rays in the atmosphere.
Higher balloon float altitudes are achieved by reducing the total mass of the balloon payload.
Furthermore, due to the limited torque available from the WASP actuator motor system, the X-Calibur center of gravity has to be balanced to within \SI{0.08}{\mm} of the gimbal axis intersection point.
The X-Calibur telescope was balanced during the design phase by adjusting the lengths of the two truss halves according to the weight distribution on each end of the truss.
To achieve the precise balance along all three axes required for flight, only about \SI{35}{\lbs} of brass and aluminum weights had to be mounted to the bulkheads during integration.

\section{Design of the truss}\label{sec:design}
\subsection{Mechanical design}\label{sub:design:mechanical}
\begin{figure}
  \centering
  \includegraphics[width=.5\textwidth]{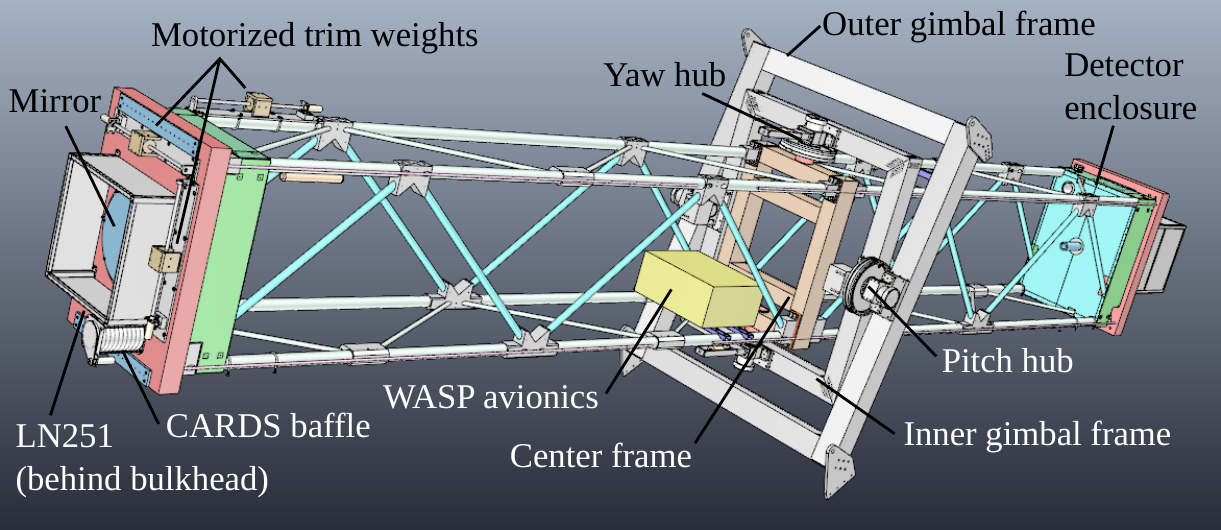}
  \caption{CAD rendering of the truss structure connected to the WASP gimbal frames. All equipment is shown mounted on the truss, and labels indicate several key components.}
  \label{fig:truss-schematic}
\end{figure}

The truss for the X-Calibur telescope, which has been designed by Guarino Engineering Services\footnote{Guarino Engineering Services, 1134 S. Scoville Ave, Oak Park, IL 60304}, is made of carbon fiber tubes and machined aluminum joints.
It consists of two halves that are bolted onto a welded and machined aluminum center frame structure, which mounts onto the yaw hubs of the WASP system.
Aluminum honeycomb panels holding the equipment are bolted to each end of the truss.
Figure~\ref{fig:truss-schematic} shows a CAD rendering of the truss attached to the WASP gimabel frames with all equipment mounted.
Key components on the front bulkhead (``mirror bulkhead'') are the InFOCuS X-ray mirror, the CARDS star tracker, and the LN251 IMU.
The WASP avionics deck including all WASP flight computers and a GPS receiver is mounted on the truss close to the center frame.
The rear, or ``detector'', bulkhead holds the detector inside its rotating CsI shield including the data taking CPU, and one of the X-Calibur flight computers.

Three types of carbon fiber tubes are used in different roles in the structure:
\begin{itemlist}
 \item the four main load-carrying cords have an outer diameter of \SI{2.5}{"} and a wall thickness of \SI{0.25}{"};
 \item tubes with an outer diameter of \SI{1.5}{"} and a wall thickness of \SI{0.25}{"} are used for the side diagonal braces;
 \item the top and bottom lateral diagonals are made of \SI{1.0}{"} diameter tubes with a wall thickness of \SI{0.125}{"}.
\end{itemlist}
Commonly used carbon fiber tubes have a wall thickness of \SI{0.125}{"}.
Using thicker-wall tubes increases their stiffness and allows using smaller diameter tubes.
This reduces the required joint size and hence the truss mass, as well as the thermal expansion of the joints, and hence thermal stability of the truss.
The tubes are glued to the aluminum joints using Loctite Hysol E-120HP epoxy.
Hence, each of the two truss halves is a single rigid-glued part.
This construction is stiffer than other approaches in which aluminum end pieces glued to the carbon fiber tubes are bolted to joints.
The downside is that the glued truss structure is harder to repair.
There are three types of carbon fiber-to-aluminum glue joints.
All diagonals slide into a hole in the aluminum blocks (``female'' joints).
This type of joint turned out to be strongest, in particular after exposing the joint to cold temperatures (see Section~\ref{sub:component-tests}).
However, to reduce the size of the aluminum pieces, a different approach was chosen for the main cords, where an aluminum piece slides into the tube (``male'' joints).
In three places on each half truss there are ``slide-through'' joints, where an aluminum joint is slid onto a through-going main cord.

In order to eliminate first-order thermal bending of the truss, it was ensured that there is an equal amount of aluminum within each cord.
Note that the slide-through joints were not considered here, since they do not contribute to the thermal expansion of the cord.
This resulted in the double-length joints that can be seen in the center of the bottom cords in Fig.~\ref{fig:truss-schematic}.

The carbon fiber tubes were ground on the outside by the manufacturer in order to achieve a tolerance better than \SI{0.002}{"} on the outside diameters.
The joints were machined to achieve an \SI{0.008}{"} bond line with a precision of \SI{0.002}{"}.
The inside surfaces were less precise and in some cases bumpy, requiring additional grinding.
The resulting variations of the inside diameters made it necessary to adjust all male joints in order to achieve a bond line thickness of \SI{0.012}{"} to \SI{0.015}{"}.
All female and slide-through joints were honed, and all male joints were bead-blasted after machining in order to roughen the surfaces.
Finally, all joints were chemically etched in order to remove the oxidation layer, and cleaned with Acetone.
For glueing, the epoxy was mixed with \SI{2}{\percent} by weight of \SI{0.007}{"} and \SI{0.012}{"} glass beads for the female and male joints, respectively.

\begin{figure}
  \centering
  \includegraphics[width=.5\textwidth]{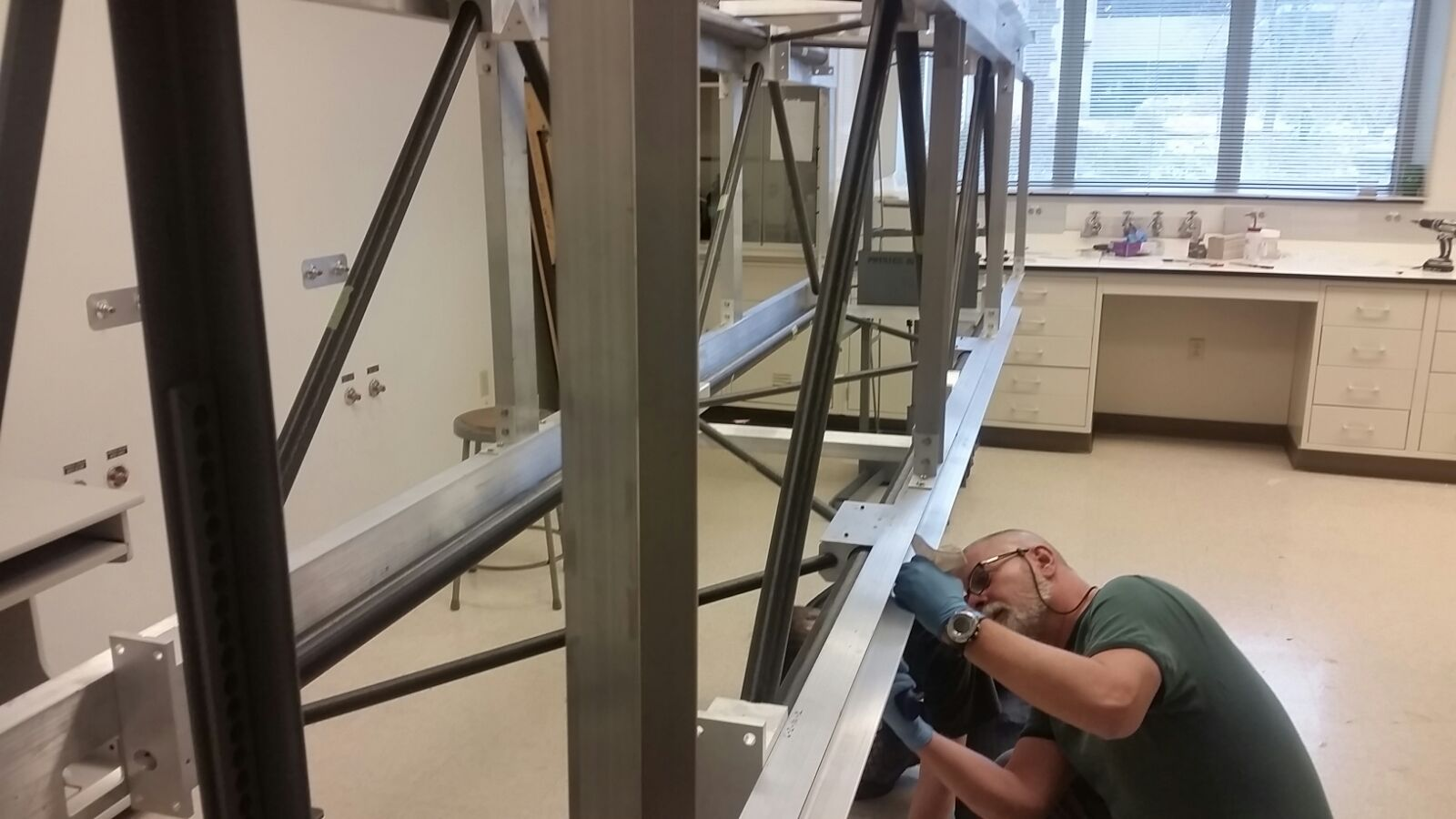}
  \caption{Truss assembly structure during the glueing of the rear half of the truss. Joints were bolted to the structure during assembly and curing in order to ensure precise alignment of all elements.}
  \label{fig:assembly-structure}
\end{figure}

The truss was assembled using the structure shown in Fig.~\ref{fig:assembly-structure} with all joints bolted to the structure to fixate them during assembly and for curing.
This structure was designed such that it could accommodate both truss halves, despite their different length.
When gluing the male joints, both surfaces were coated with epoxy.
In order to glue the female and slide-through joints, epoxy was injected with a syringe through several small holes located all around the joint.
These approaches ensured an even coating of all surfaces, as our tests showed (see Section~\ref{sub:component-tests}).

The honeycomb panels are 3/8" vented aluminum honeycomb with a thickness of \SI{4}{"} and \SI{0.020}{"} thick face sheets.
The edges of the panels are filled with epoxy to protect the thin aluminum honeycomb.
Equipment is mounted to the panels using threaded inserts epoxied into holes in the honeycomb with Loctite Hysol 9394 epoxy.

For thermal control the entire truss was painted with white appliance epoxy paint, then wrapped in fine mesh, and finally wrapped in \SI{0.005}{"} thick aluminized mylar.
The purpose of the mesh is to prevent thermal contact between the mylar and the truss structure.
The honeycomb panels and all other exposed surfaces were painted white.
On the mylar-covered surfaces of the truss the white paint mainly served safety purposes: should a piece of mylar be lost, for instance during launch, no black carbon fiber or unprotected aluminum surfaces would be exposed to the sun.

In addition to designing the truss in a way that it would remain straight despite temperature variations during the flight, we also added heaters and temperature sensors to each joint.
Both manual control and an automatic algorithm to keep the joints at equal temperatures were implemented.
However, the system was not used during the September 2016 flight, since all focal spot offsets were within an acceptable range.
In part, this can be attributed to the fact that the temperature gradients along the truss were significantly lower than the predicted extremes for most of the flight.

\subsection{Alignment monitoring systems}\label{sub:design:monitoring}
We employed two systems to monitor the alignment between mirror focal position and detector during the flight.
The first one, which has already been used on the first X-Calibur flight with the InFOCuS telescope, is a camera mounted to the central bore-hole of the mirror, observing the detector.
Two rings of 8 LEDs each are mounted on a black circuit board around the entrance window of the detector showing up as bright spots in the image taken with the camera.
The advantage of this arrangement is that both parts are centered on the optical axis of the mirror, and all observed translations directly correspond to deviations of the focal spot from the detector center.
One pixel in an image of this LED ring corresponds to a deviation of \SI{0.5}{\mm}, and by observing multiple LEDs each extending across several pixels, an accuracy of the focal spot location of \SI{0.1}{mm} was achieved.

During the first flight, the LED ring was mounted to the outside of the X-Calibur pressure vessel.
An automatic algorithm to find the LED locations was used on the flight system.
All pixels within a configurable radius around the last known position of each LED were inspected to see if they are above a programmable threshold.
Clusters of pixels above threshold were then constructed using a breadth-first traversal of the image starting from an above-threshold pixel.
The center of each LED in the image was then calculated as the weighted average location of all pixels above threshold in the largest such cluster.
On the ground, the known pattern of LED locations was then fit to the 16 cluster locations, resulting in the coordinates $x$, and $y$, the image scale $\rho$, and the rotation angle $\alpha$.

During the September 2016 flight, the LED ring was mounted directly to the rotating CsI shield.
The algorithm to detect the rotating LED positions was not completed in time for the flight.
Instead, all images were stored on the flight system and analyzed manually in the post-flight analysis.

\begin{figure}
  \centering
  \includegraphics[width=.4\textwidth]{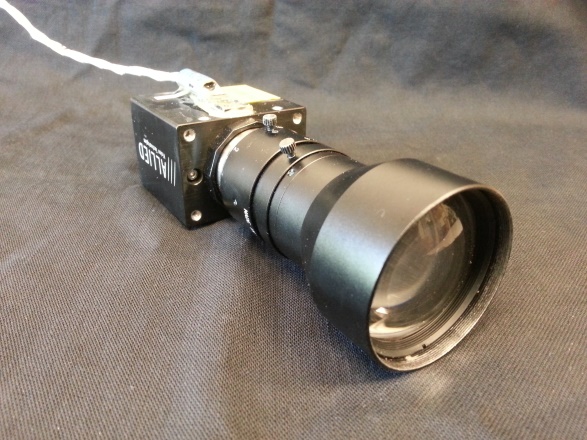}\quad\includegraphics[width=.4\textwidth]{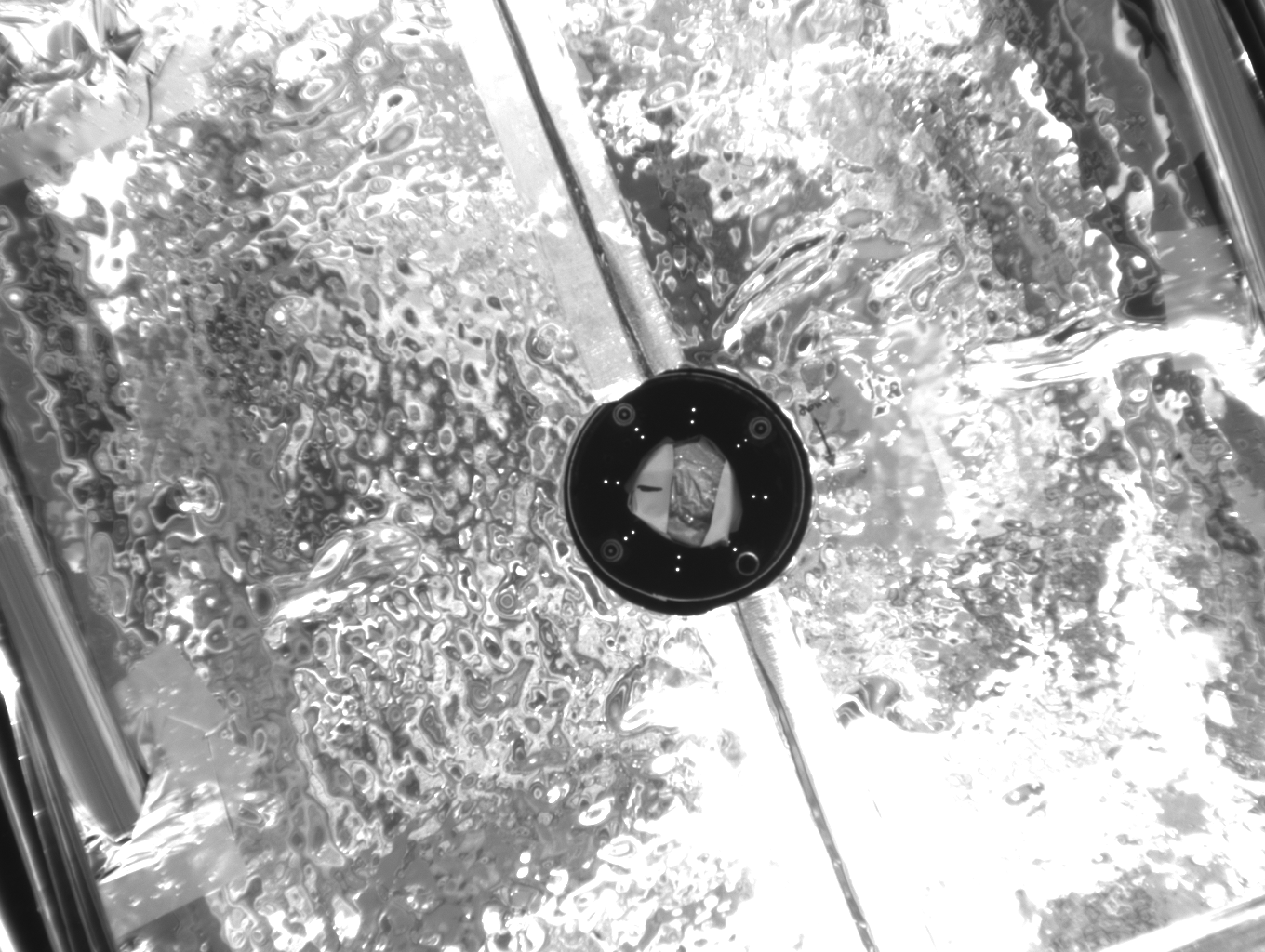}
  \caption{\emph{Left:} The backlooking camera is a Allied Vision Prosilica GC 1380 equipped with a \SI{100}{\mm} telephoto lens. \emph{Right:} A typical image observed with the backlooking camera during the flight. The black disk populated with 16 LEDs is mounted directly to the top of the active shield. The camera is upside-down and rotated slightly with respect to the vertical due to mounting constraints on the X-ray mirror which needs to be taken into account when analyzing the images.}
  \label{fig:backlooking-image}
\end{figure}

The second alignment monitoring system consists of a \SI{405}{\nm} wavelength laser diode collimated using an aspheric lens with an effective focal length of \SI{4.02}{mm} resulting in an elliptical beam with a FWHM major axis diameter of about \SI{1.4}{\mm}.
The beam is directed at a $244 \times 550$ pixel CCD with an area of $8.8 \times \SI{6.6}{\mm}$ with a custom-built readout system.
This system is a space-qualified spare from the development of the CRIS instrument~\citep{1998SSRv...86..285S} on the Advanced Composition Explorer, chosen because of its known flight qualification.
The brightness of the laser beam was adjusted to match the CCD sensitivity using a neutral density filter.
With its $\SI{27}{\um} \times \SI{16}{\um}$ pixel pitch this system offers a significantly better spatial resolution than the back-looking camera.
However, space constraints resulted in the system being mounted significantly away from the optical axis of the X-ray mirror meaning that it will not only be sensitive to truss bending but also to twisting deformations.
The readout system of the CCD camera is designed to only report pixels above an adjustable threshold.
This threshold was tuned before the flight and a simple weighted average of the pixels above threshold was calculated by the flight system and reported to the ground as the location of the laser spot.

For diagnostic purposes we implemented the ability to download images of both alignment systems from the payload during flight.
In case of the back-looking camera, images were compressed using the GIF format and optionally cropped in order to reduce the file size.

\section{Structural and thermal analysis}\label{sec:analysis}
\begin{figure}
 \centering
 \includegraphics[width=0.4\textwidth]{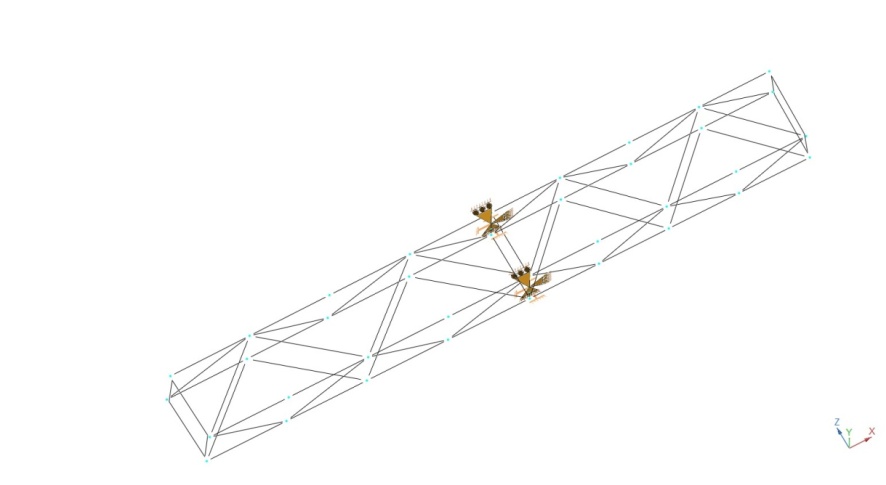}
 \caption{Simple beam and plate finite element model of the truss used during the early development in order to determine the expected loads and deformations during flight.}
 \label{fig:beam-plate-model}
\end{figure}

Initial dimensioning of the truss members was based upon simple empirical calculations, assuming the following carbon fiber properties given by the manufacturer, CST Composites:
\begin{itemlist}
 \item Axial modulus of elasticity: \SI{14}{Msi};
 \item Ultimate tensile strength: \SI{140000}{psi};
 \item Density: \SI{0.06}{\lbs/\inch^3};
 \item Coefficient of thermal expansion: \SI{0.18e-6}{\kelvin^{-1}};
\end{itemlist}
A simple beam and plate element finite element model was then created, see Fig.~\ref{fig:beam-plate-model}.
The maximum axial tensile force in the chords is \SI{682}{\lbs} and the maximum diagonal member tensile force is \SI{346}{\lbs} when the truss is horizontal.
The tensile tests showed that the joints can achieve significantly greater strength than what is required (see Section~\ref{sub:component-tests}).  

\begin{figure}
 \centering%
 \includegraphics[width=0.49\textwidth]{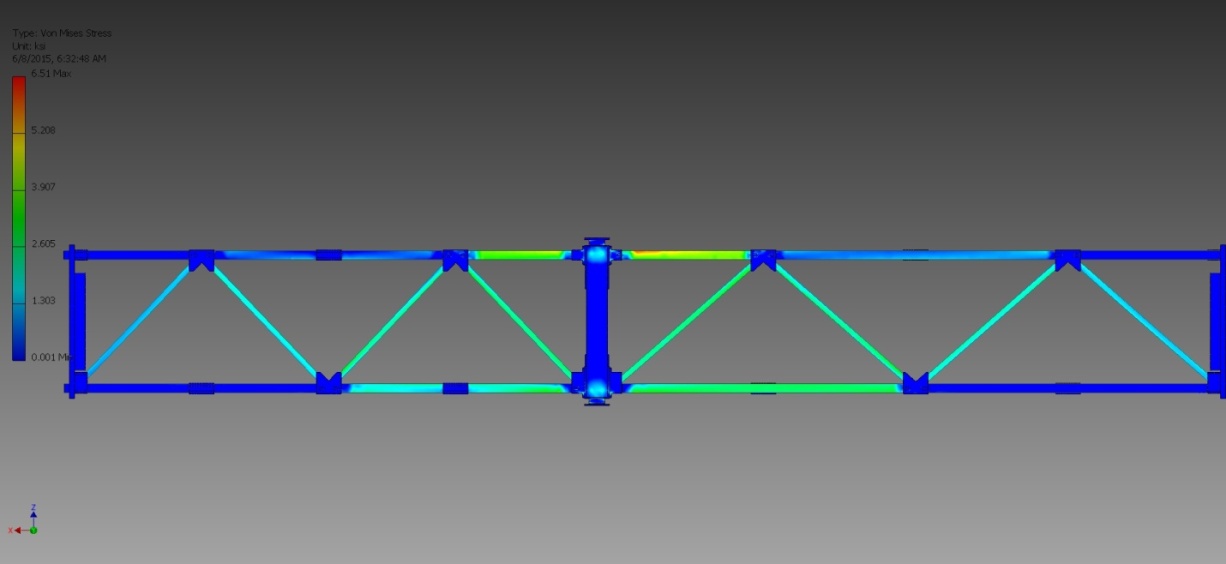}\hfill\includegraphics[width=0.49\textwidth]{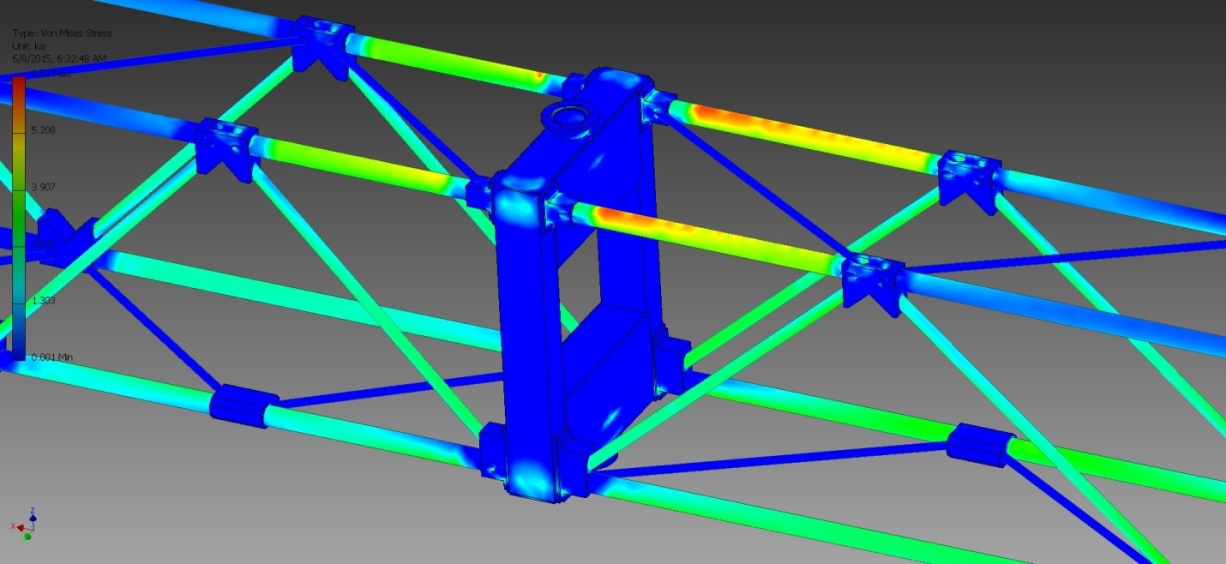}\\
 \includegraphics[width=0.49\textwidth]{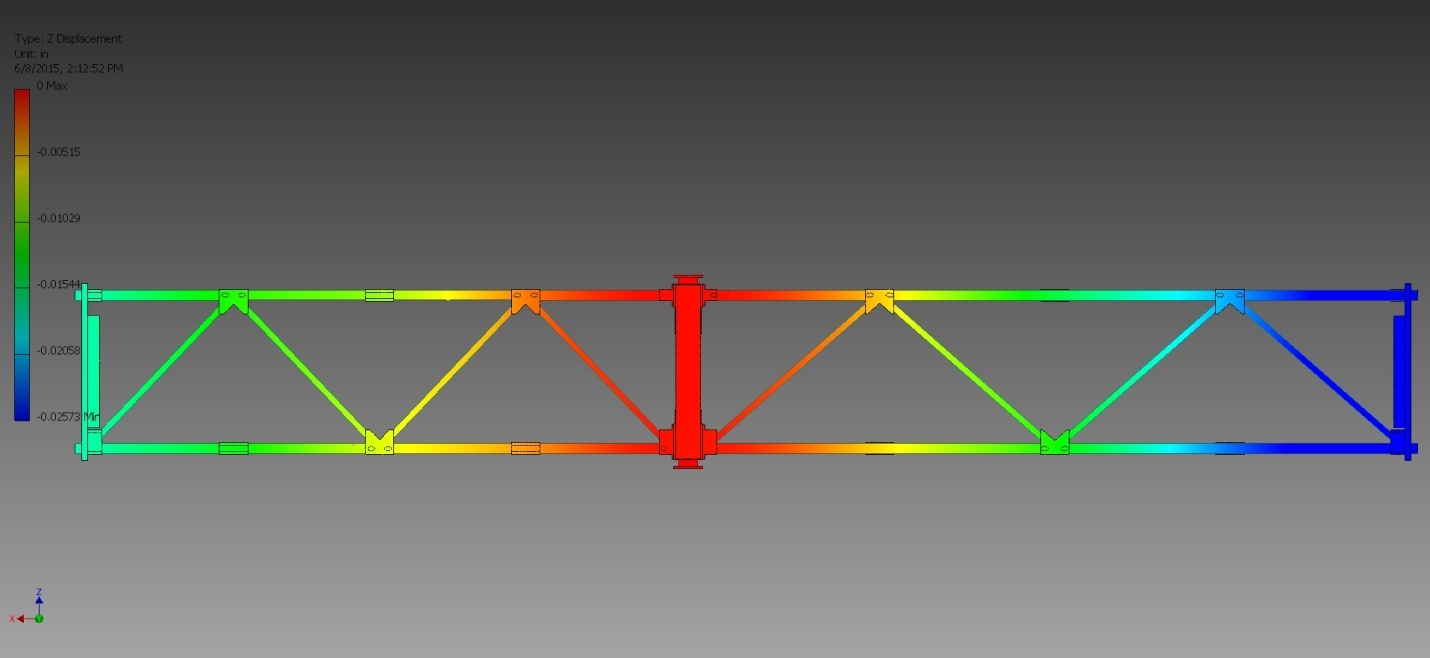}\hfill\includegraphics[width=0.49\textwidth]{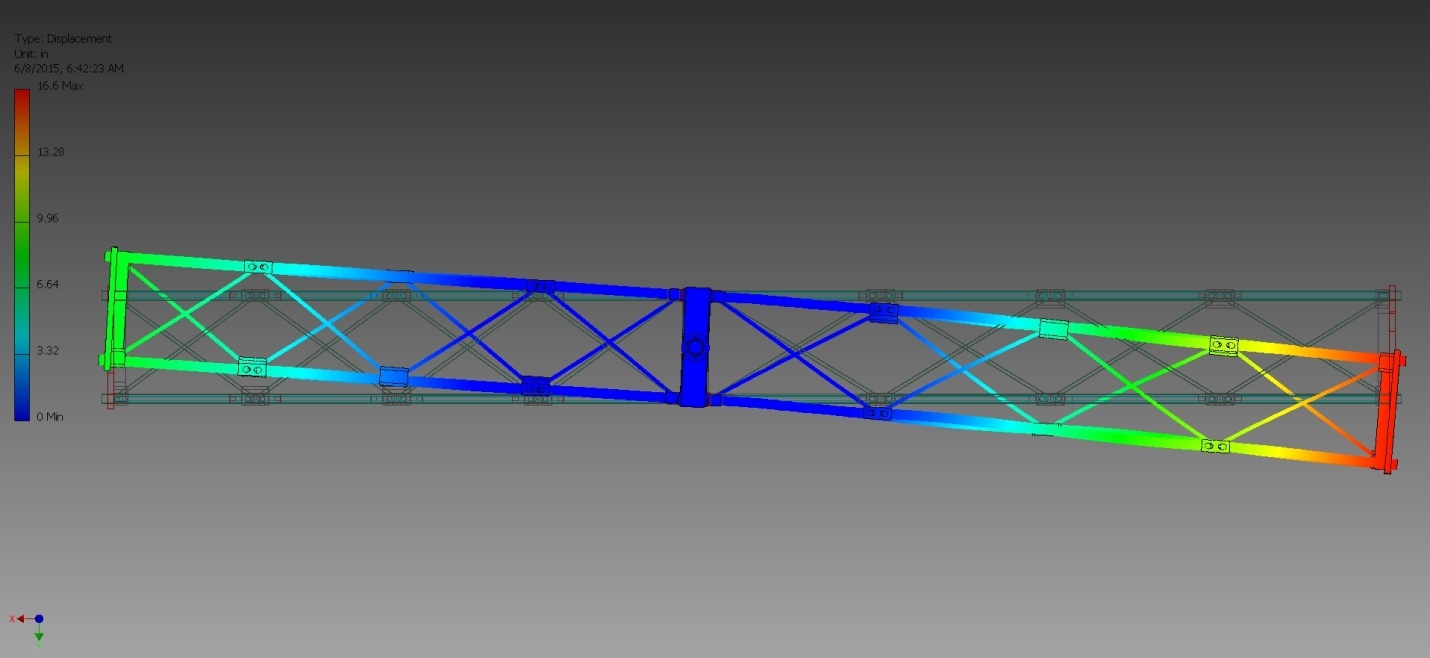}%
 \caption{\emph{Top row:} Stresses on the truss members, when the truss is oriented horizontally as during launch and flight termination, assuming a $10g$ load. \emph{Bottom left:} Deformations of the truss when it is oriented horizontally, assuming a $1g$ load. Locations of the honeycomb mounting points were used to calculate the orientation of the mirror and detector, and thus the offset of the X-ray beam from the detector center. Equivalent calculations were performed at truss elevations of \SI{20}{\degree}, \SI{45}{\degree}, and \SI{60}{\degree}. \emph{Bottom right:} Modal shape of the first natural frequency of the truss at \SI{10.4}{Hz}.}
 \label{fig:fea-results}
\end{figure}

Finally, a solid element model of the complete structure was created in order to evaluate expected deflections under load as well as its natural frequencies.
Examples of the resulting stresses and deformations are shown in Fig.~\ref{fig:fea-results}.
It should be noted that the axial load on the truss members agreed in all three approaches (empirical, beam element FEA, and solid element FEA).
Even at $10g$ vertical load, stresses on all truss members are significantly below the ultimate tensile strength of \SI{140000}{psi}.
We also evaluated the system at a $5g$ horizontal load, as required by the guidelines for gondola design set forth by NASA's Columbia Scientific Balloon Facility (CSBF), and found maximum stresses of \SI{7700}{psi}.

The center frame that the carbon fibers trusses are bolted to are made of welded aluminum 6061T6 extruded tubes.
The area of the welded heat affected zones has a de-rated yield strength of \SI{11000}{psi}.
The stresses in the weld areas of the extreme survival load cases of $10g$ vertical and $5g$ lateral loading showed stresses of less than \SI{1000}{psi}.

Based on the design of the truss and gondola, a Thermal Desktop$^\circledR$ model was developed by Scott Cannon at the New Mexico State University Physical Science Laboratory, including the suspension cables and the balloon, as well as all sources of heat.
Two back-to-back 24-hour transient analyses of radiation and thermal flow were then performed using RadCAD$^\circledR$ and SINDA$^\circledR$.
The first 24-hour period was initialized from an orbital average configuration and used to allow temperatures to settle.
The second 24-hour period was initialized from the results of the first run and used to predict the flight temperature extremes.
Simulations were performed for four scenarios:
\begin{itemlist}
 \item A Ft.\ Sumner, NM, hot case assuming an afternoon over the cloudless hot desert;
 \item A cold case at night in Ft.\ Sumner, just prior to sunrise, with a cold convective cloud cover;
 \item A hot case at McMurdo, Antarctica, assuming a typical maxima in solar and earth flux as well as albedo, approximately at noon local time near the December solstice;
 \item A cold case at McMurdo approximately at midnight local time, close to the end of the flight window on January 31st.
\end{itemlist}
Predictions were taken as the most extreme temperature anywhere on each component, any time during the 24-hour cycle.
These predictions were then compared to the required temperature ranges and passive thermal control and required heating were adjusted accordingly.

\begin{figure}
 \centering
 \includegraphics[width=0.6\textwidth]{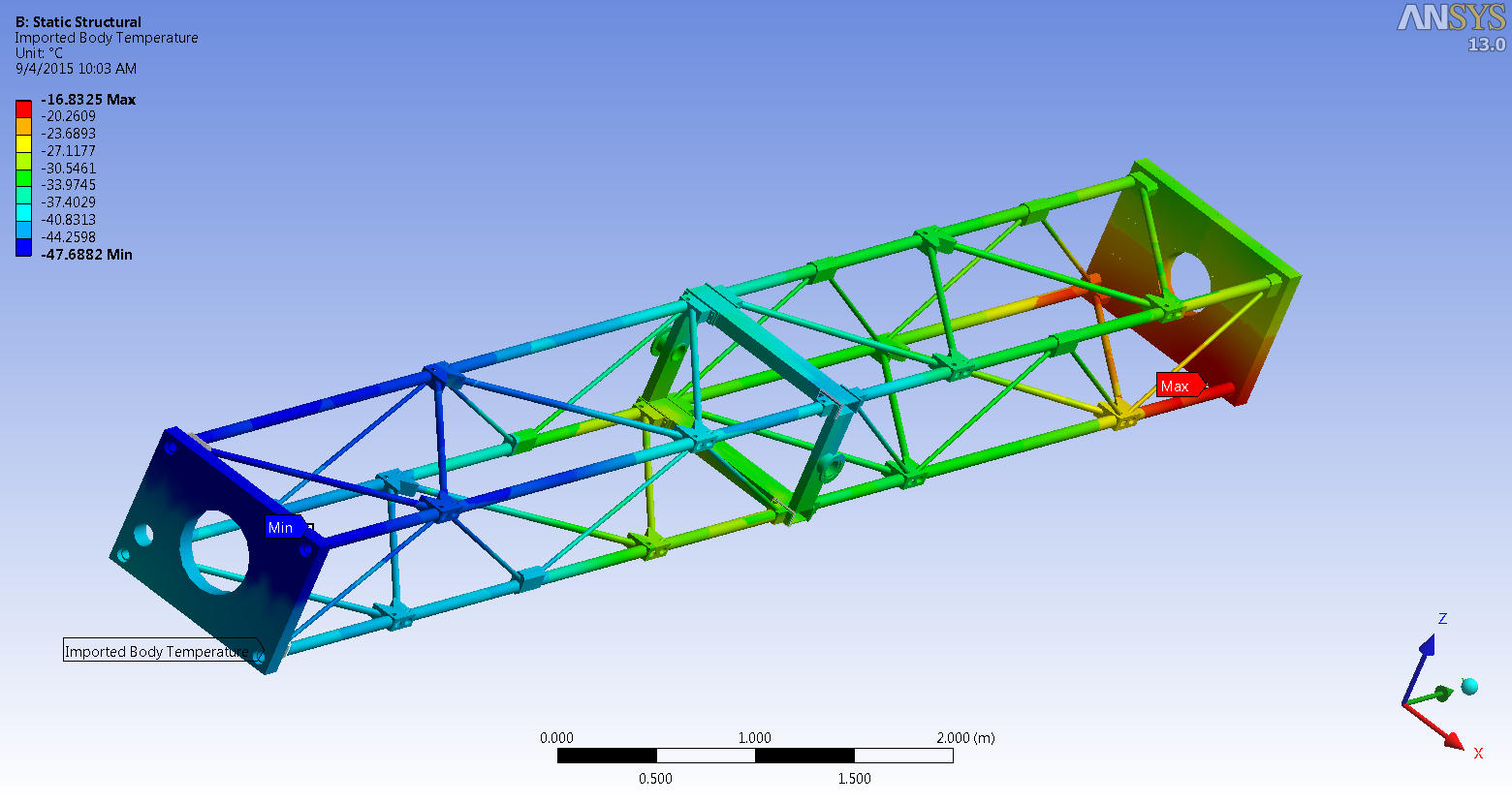}
 \caption{Extreme example of a predicted temperature distribution along the X-Calibur truss. These temperature predictions were then used to calculate mechanical deformations of the truss due to the change from room temperature, at which the X-ray mirror was aligned with the detector (see Section~\ref{sub:requirements:pointing}).}
 \label{fig:truss-temperature-prediction}
\end{figure}

The resulting temperature predictions (for an example see Fig.~\ref{fig:truss-temperature-prediction}) were then incorporated in the FEA model in order to predict thermal deformations of the truss.
One of the results of this modeling was the decision to increase the size of the joints in the lower truss cords, such that there is an equal length of aluminum in each cord (see Section~\ref{sec:design}).

\section{Tests and verification}\label{sec:tests}
\subsection{Component tests during the design phase}\label{sub:component-tests}
The design phase was accompanied by a series of tests in order to ensure the final product would meet the design requirements.
The focus of these tests were the glued connections between the carbon fiber tubes and the aluminum joints, as well as the inserts in the honeycomb panels.

The strength of the glued joint connections is determined amongst others by the following factors:
\begin{itemlist}
 \item the type of glue used;
 \item surface preparation;
 \item thickness of the bond line;
 \item size of the glued area (and covering fraction of the glue);
 \item and conditions during curing, in particular the stability of the assembly.
\end{itemlist}

The flight on a stratospheric balloon will expose the telescope truss to a wide range of temperatures, ranging from \SI{-60}{\celsius} at night up to \SI{+50}{\celsius} inside the container when the truss is shipped to the field.
The glues used in the truss assembly must be able to maintain their strength throughout this temperature range.
Furthermore, CSBF requires that gondolas are designed to withstand a $10g$ shock, which can occur during flight termination when the parachute opens.
We performed a series of pull tests of aluminum joint samples in order to determine the best procedures.

\begin{figure}
  \centering
  \includegraphics[height=6cm]{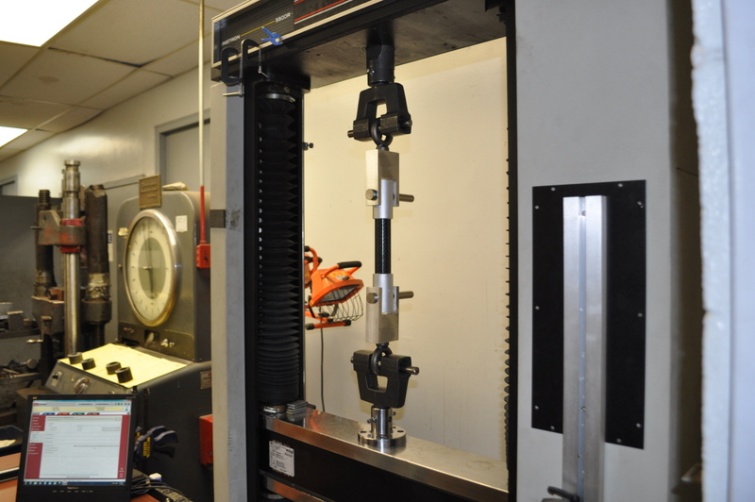}\quad\includegraphics[height=6cm]{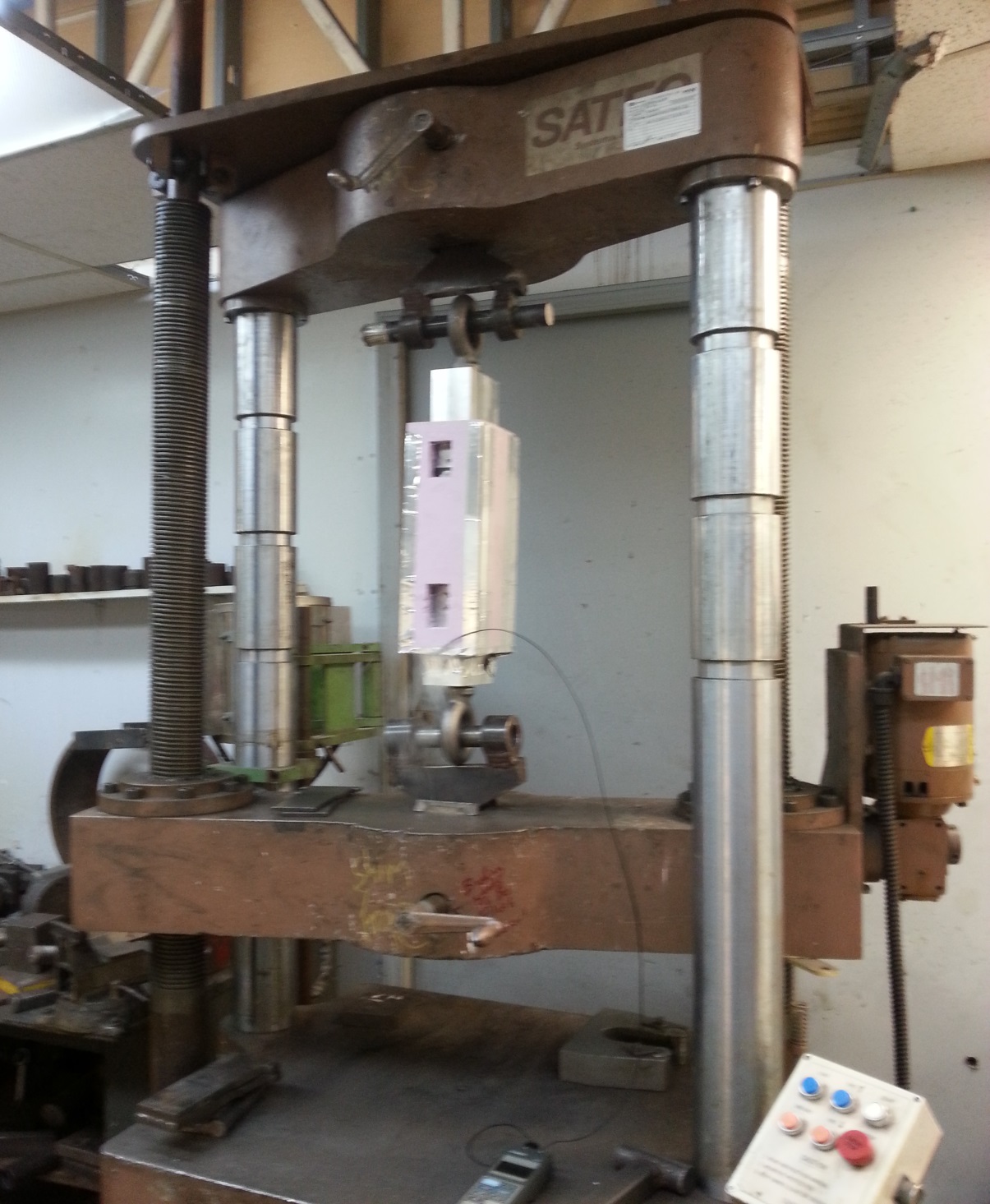}
  \caption{Setup to pull test the carbon fiber-aluminum joint samples. All joints were pulled to failure in order to determine the ultimate strength. \emph{Left:} Pull-testing small samples of the first batch at room temperature. \emph{Right:} Test of a flight-sized sample embedded in dry ice.}
  \label{fig:pull-test}
\end{figure}

In a first set of tests, three different types of adhesive (3M 2216, 3M DP-460, and Loctite Hysol E-120HP) were compared.
Different methods of surface roughening (sanding, bead blasting, honing female joints, grooving) and surface treatment (etching, anodizing, iriditing) were tested.
Furthermore, different tube diameters and glue lengths were compared.
The samples were pull-tested at Saint Louis Testing Laboratories, Inc.\ with the setup shown in Fig.~\ref{fig:pull-test}, in which all samples were pulled to failure.
The results of these tests were used to choose the final configuration described in Section~\ref{sub:design:mechanical} and to decide on the lengths of the individual joints based on the expected stresses.
In a second set of tests, flight-sized joints were prepared with the previously selected methods.
Before pull-testing these samples, they were exposed to the temperatures expected during shipping and flight.
Four samples of each of the flight-size joints were tested:
\begin{itemlist}
 \item 1 sample without temperature cycling;
 \item 1 sample that went through half of the temperature cycling;
 \item 1 sample that went through the full temperature cycling procedure (see below);
 \item 1 sample that went through the full temperature cycling and was kept at roughly \SI{-78}{\celsius} during the pull test using dry ice.
\end{itemlist}

The temperature cycles were:
\begin{arabiclist}[(3)]
 \item 8 cycles \SI{+20}{\celsius} from to \SI{+55}{\celsius} to simulate shipping container temperatures;
 \item 3 cycles from \SI{0}{\celsius} to \SI{-65}{\celsius} to simulate Ft.\ Sumner flight and trope conditions;
 \item 8 cycles from \SI{+55}{\celsius} to \SI{-45}{\celsius} to simulate shipping and Antarctic flight temperatures.
\end{arabiclist}
Each temperature set point was maintained for a 2-hour soak period with \SI{2}{\celsius/min} ramps in between.
Additionally, there was one accidental about 24 hour long soak at \SI{-70}{\celsius} due to a failure of the temperature controller.
The sample embedded in dry ice during the test was the weakest and failed a load of \SI{16000}{\lbs}.
Maximum expected $10g$ loads on the main cords are \SI{6820}{\lbs}.

\begin{figure}
 \centering
 \includegraphics[width=.3\textwidth]{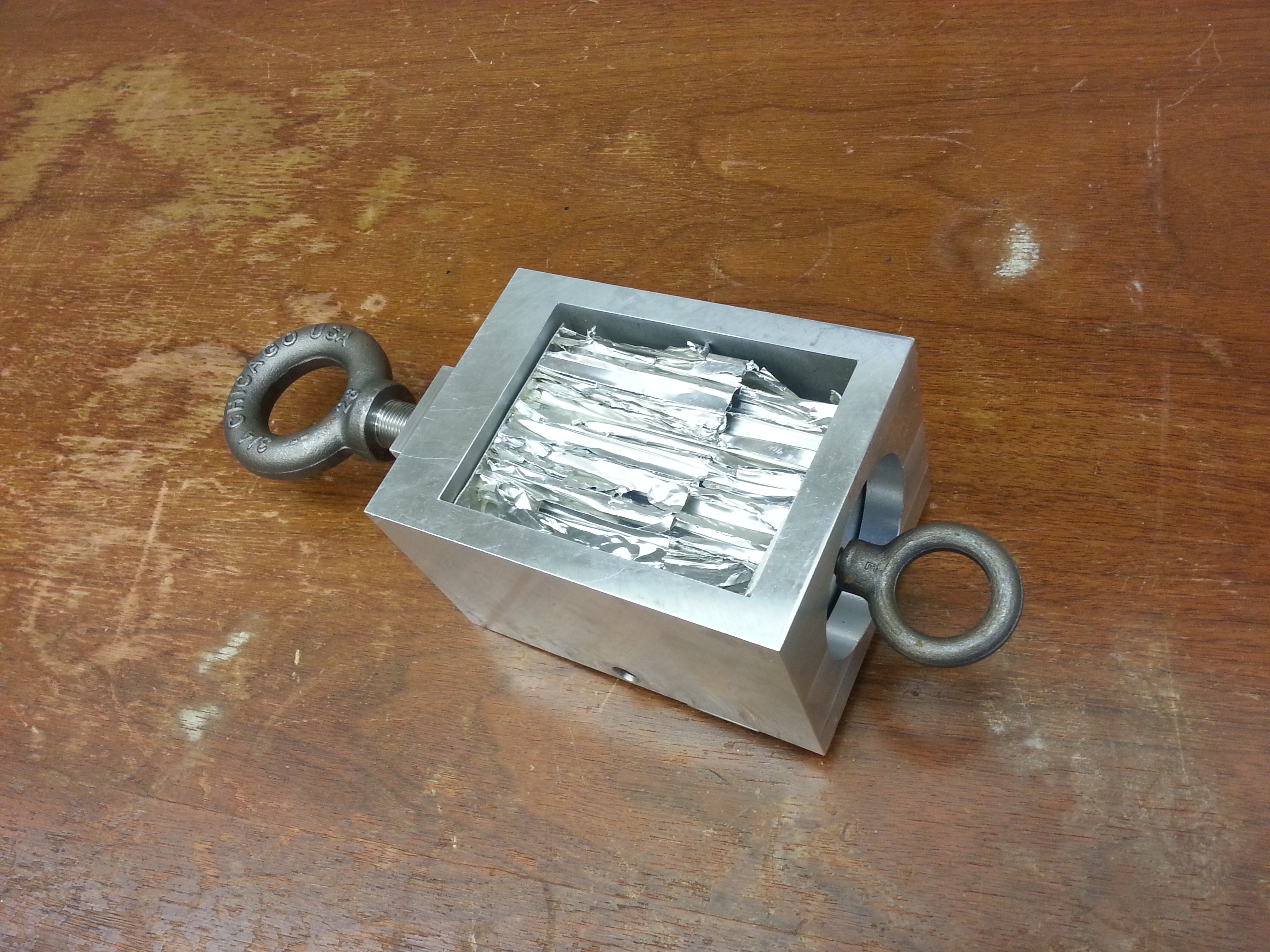}\quad\includegraphics[width=.3\textwidth]{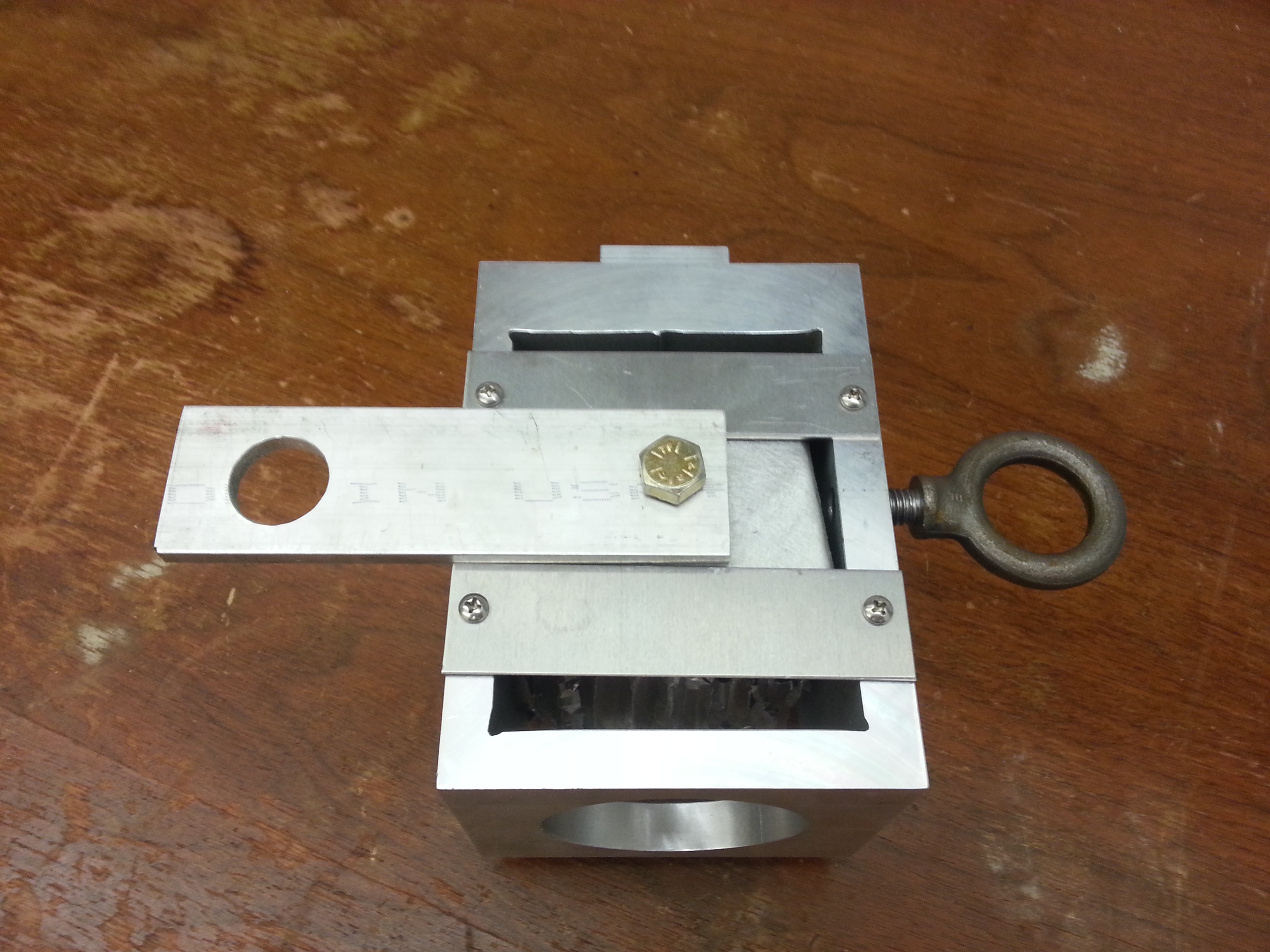}
 \caption{Honeycomb pull and shear test samples. The test coupon was held in the fixture as shown while pull tested on the same machine that was used for the carbon fiber joint tests. \emph{Left:} Setup for pull tests. \emph{Right:} Shear test configuration.}
 \label{fig:honeycomb-test}
\end{figure}

The honeycomb panels were manufactured by Kerr Panel Manufacturing, Inc.
Test samples of the honeycomb panel and glued fastener inserts were pull and shear tested under similar conditions to those of the joint tests described above.
Tensile strength testing was performed on the same machine that was used to test the carbon fiber samples.
Honeycomb test coupons with glued inserts for 3/8" and 1/4" threads were held in the fixture shown in Fig.~\ref{fig:honeycomb-test}.
An issue during shear tests was that the face sheets of the honeycomb started to fold at the edges before failing at the insert.
Hence, the results only presented a lower limit on the strength of the inserts.
These lower limits already exceeded the required $10g$ shear strength of \SI{188}{\lbs} by more than a factor of 5.

We also tested different methods of glue application by glueing clear plastic tubes to cylindrical aluminum pieces.
We found that in particular in case of the female and slide-through joints where it is impractical to cover both surfaces, too much glue might be scraped off.
Injecting the glue through holes on four sides until glue can be seen leaking out at the end of a joint reliably covered the entire surface.

\subsection{Verification of the assembled truss}
\begin{figure}
  \centering
  \includegraphics[width=.4\textwidth]{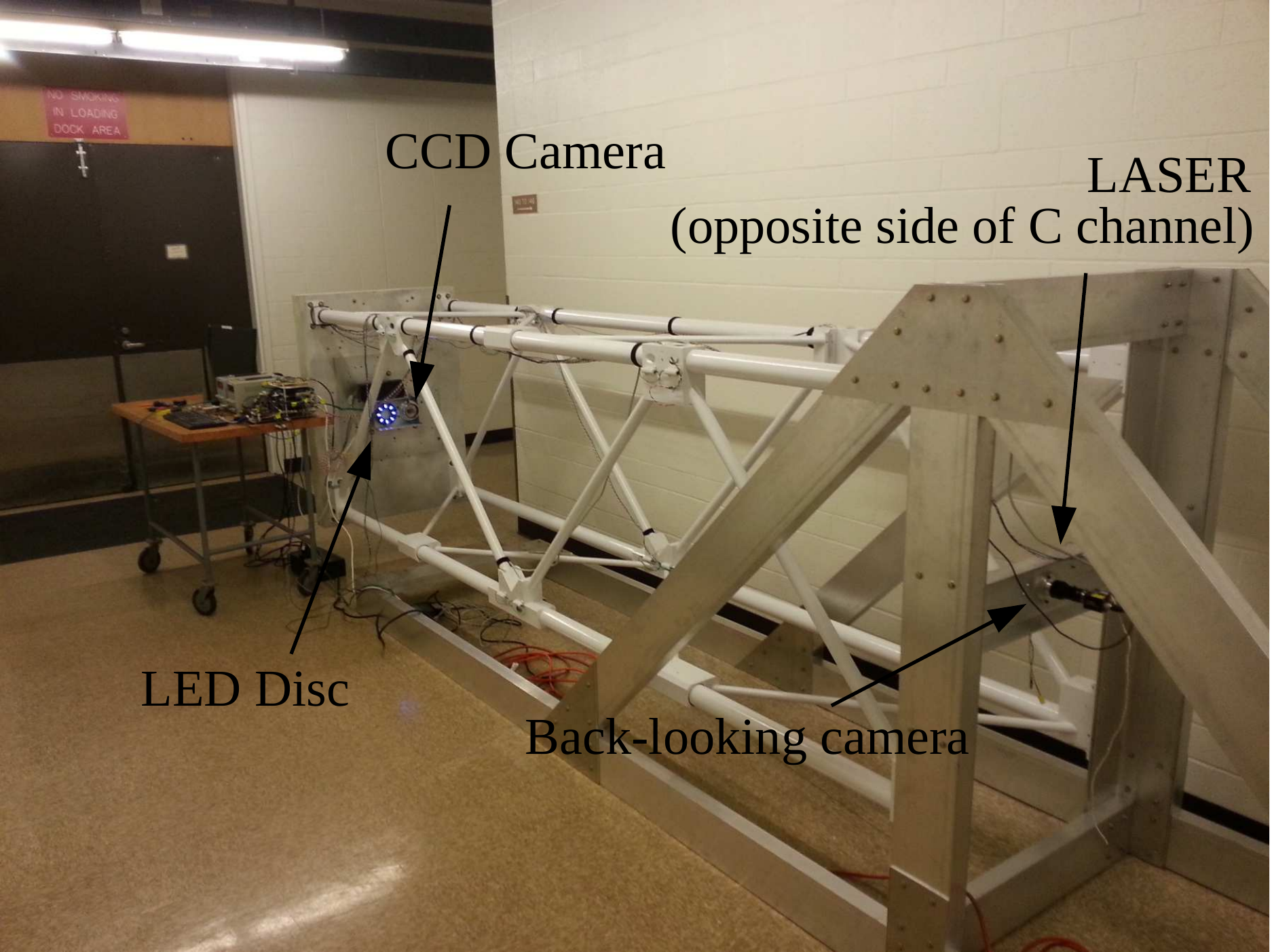}\quad\includegraphics[width=.4\textwidth]{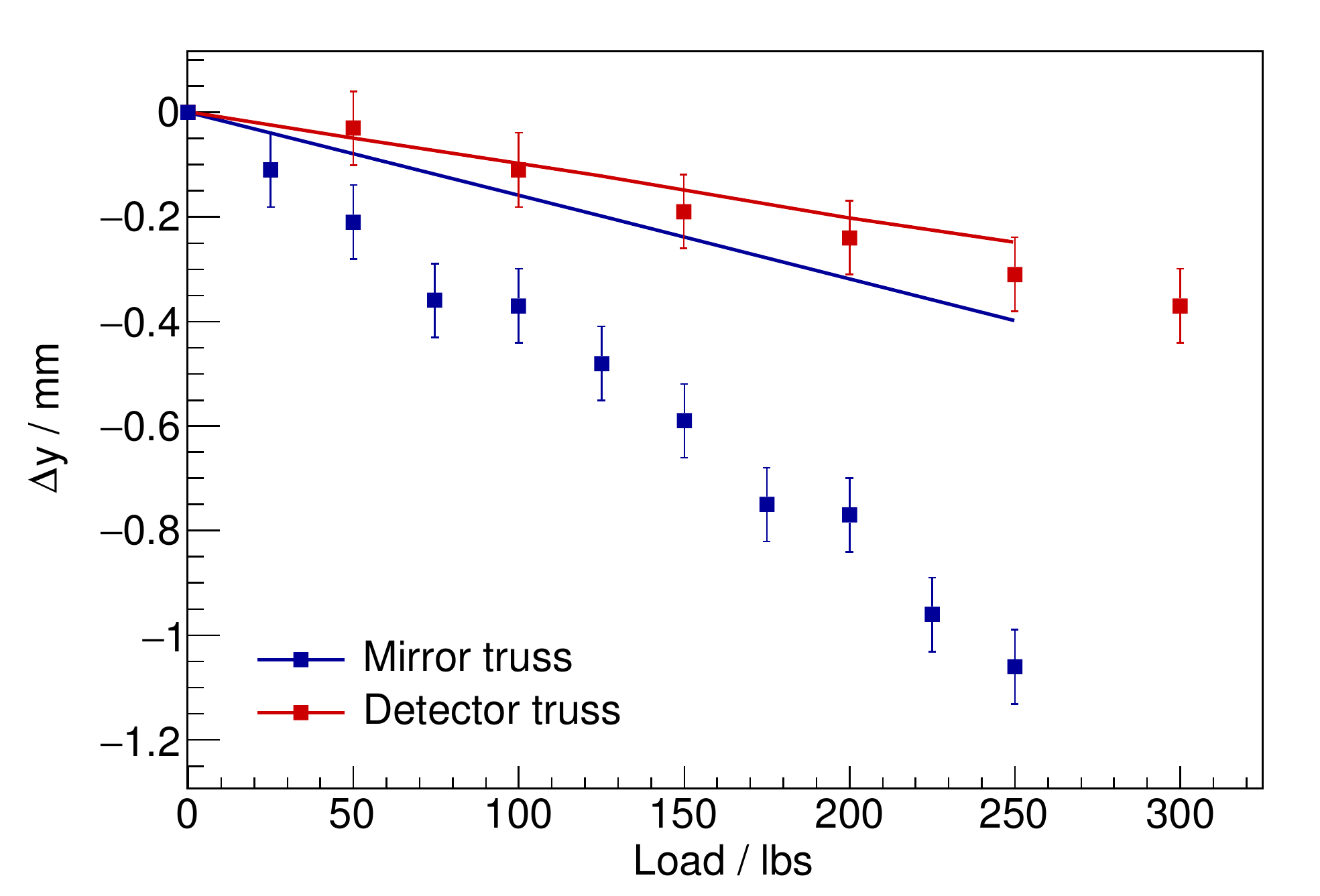}
  \caption{\emph{Left:} Load testing assembly of the read half of the truss. The truss was attached to the stage as shown with one end supported only by the truss itself. Lead weights were then loaded onto C channels mounted to the bulkhead (opposite side, not visible in picture). The flight deflection monitoring systems were used to measure deflections of the truss. \emph{Right:} Vertical deformations of the two truss halves as a function of load on the bulkhead. Data points indicate the vertical offset of the center of the LED ring from the image center of the back-looking camera, relative to its location when the truss is not loaded. Mirror and detector truss refer to the front and rear halves of the truss, respectively.}
  \label{fig:load-testing}
\end{figure}

After assembly, each truss half was tested to ensure that it would meet (a) our stiffness requirements with respect to changes of the gravitational pull, and (b) failure strength requirements.
In order to perform these tests, each completed half-truss was mounted on one end to the cantilever stiffness test structure shown in Fig.~\ref{fig:load-testing}, while the other end was left unsupported.
This free-floating end of the truss was then loaded with weights in order to simulate the changing gravitational pull as the truss is elevated during pointing.
We used the back-looking camera system described in Section~\ref{sub:design:monitoring} to measure deflections of the truss.
The LED ring was mounted on the floating bulkhead, and the camera was attached to the fixed end of the truss\footnote{Initially we mounted the camera to the deflection test stage as shown in Fig.~\ref{fig:load-testing}. However, the structure itself warped when the truss was loaded, resulting in inaccurate measurements.
Placing the C-channel that holds the camera on top of the bottom joints, which attach the truss to the stage, circumvented the issue.}.

The measured vertical deformations of the truss as a function of load are shown in the right panel of Fig.~\ref{fig:load-testing}.
They are compared to predictions derived from an finite-element analysis of the loading of the front half of the truss.
Predictions for the rear truss half were obtained by scaling the deflections with the cube of the length of the truss, i.e.\ $(L_\text{rear}/L_\text{front})^3 = 0.633$.
In case of the rear half-truss, a very good agreement between FEA prediction and measurement was observed.
For the front half of the truss, on the other hand, the deflections were a factor of $2.5$ larger than predicted, with a maximum deflection of \SI{1.06}{\mm} at a load of \SI{250}{\lbs}.
It should be noted, that these deflections are still in agreement with our requirements. 
We also measured horizontal deflections and found the following values (as seen from the center frame):
\begin{itemlist}
 \item the detector bulkhead deflected by \SI{0.18}{\mm} to the right at a load of \SI{300}{\lbs};
 \item the mirror bulkhead deflected by \SI{0.15}{\mm} also to the right at a load of \SI{250}{\lbs}.
\end{itemlist}
These deflections are likely related to slight differences in the strengths of individual truss members and were not predicted by the finite-element model.

In order to determine whether any of the deflections were permanent (i.e. not elastic), we loaded each bulkhead to its 1G flight load (\SI{250}{\lbs} on the mirror truss, \SI{300}{\lbs} on the detector truss) and then removed the weights.
This was repeated 5 times.
No permanent deformations were found on either truss half.
Prior to this test, both truss halves were loaded with \SI{500}{\lbs}, i.e. roughly twice their flight load, in order to verify their strength.
A test at 10G load was not required by CSBF and not performed in order to avoid damage to the truss prior to the flight.

\begin{figure}
  \centering
  \includegraphics[width=0.4\textwidth]{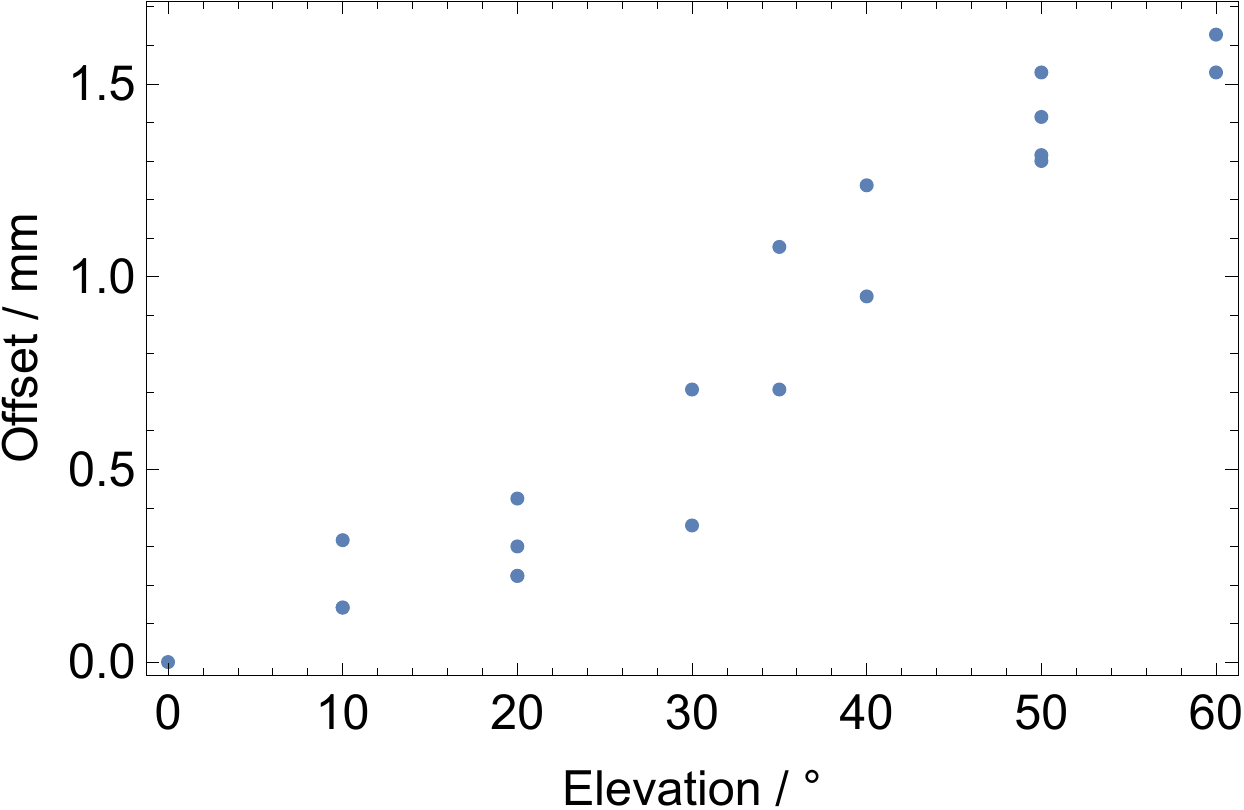}
  \caption{Truss deformations as a function of pointing elevation. Offset of the focal spot from its location in the latched position close to \SI{0}{\degree} elevation.}
  \label{fig:elevation-deflections}
\end{figure}

After integration with the WASP pointing system and gondola, we performed another set of truss stiffness measurements.
In this case we measured deflections of the pointing system as a function of elevation.
The results are shown in Fig.~\ref{fig:elevation-deflections}.
We found both vertical and lateral deformations hinting at an unexpected twist of the truss.
The results are, however, well within our limits on truss deformations.
We also observed significant hysteresis, which, however, was reset after latching the truss.

\section{Flight performance}\label{sec:flight}
Besides some minor issues, the system performed well throughout the flight.
Most importantly, the pointing system, the polarimeter, and all on-board computers worked throughout the flight.

\begin{figure}
  \centering
  \includegraphics[width=0.5\textwidth]{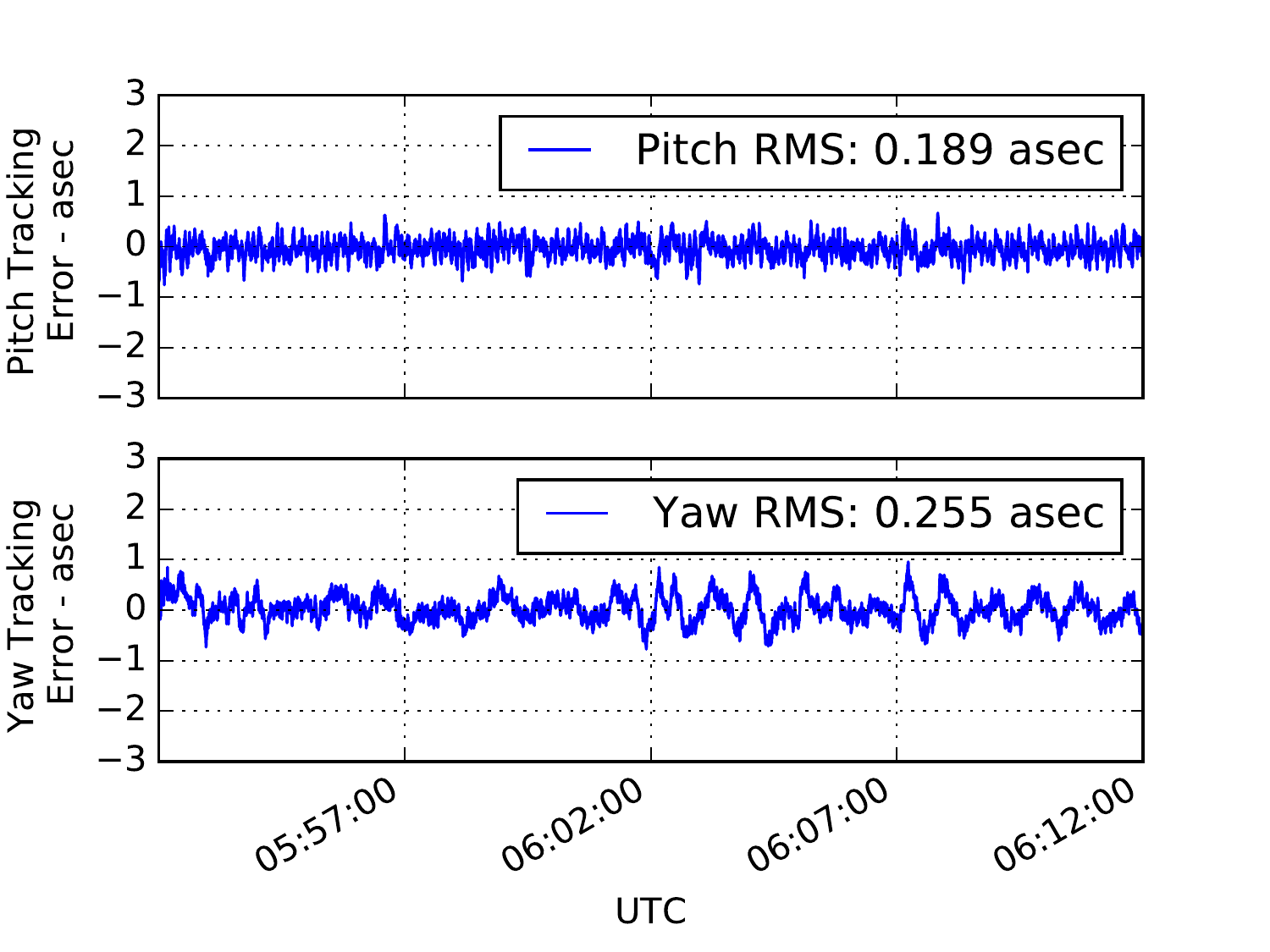}
  \caption{Plot of the tracking errors (control residuals) in the pitch and yaw axis while pointing at Cygnus X-1 during nighttime.}
  \label{fig:pointing-stability}
\end{figure}

WASP pointed with a tracking stability of \SI{0.2}{\arcsecond} RMS in pitch and \SI{0.26}{\arcsecond} RMS in yaw, exceeding the pre-flight expectations, as demonstrated in Fig.~\ref{fig:pointing-stability}.
The absolute pointing accuracy was \SI{60}{\arcsecond} during the day and \SI{15}{\arcsecond} at night.
The poorer daytime pointing resulted from a non-ideal centroiding algorithm in the star tracker software, that was influenced too strongly by background noise.
Updates to the algorithm to address these issues are currently being developed, and a pointing accuracy better than \SI{15}{\arcsecond} is expected at all times for future flights.

\begin{figure}
  \centering
  \includegraphics[height=4.5cm]{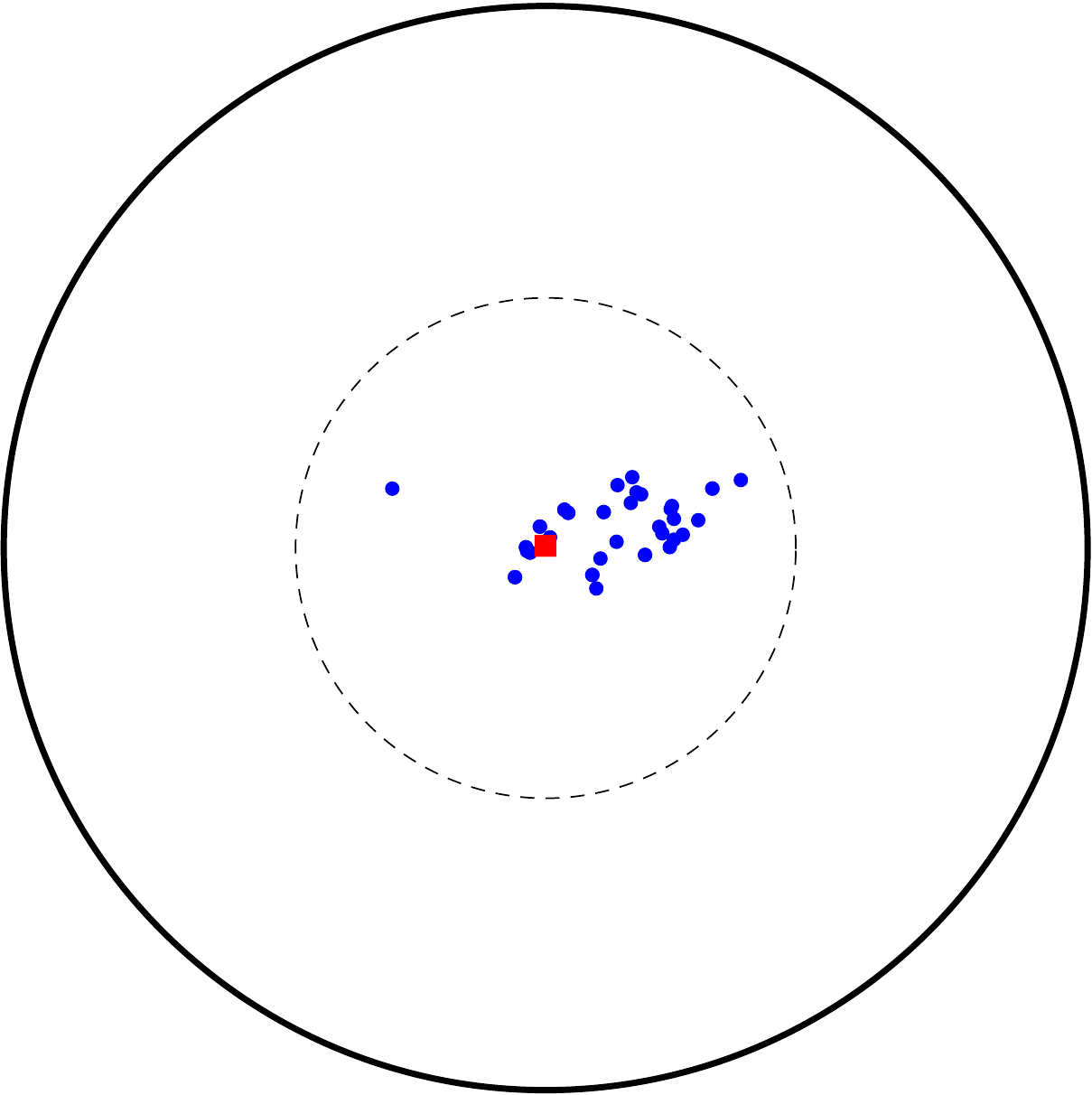}\qquad\includegraphics[height=4.5cm]{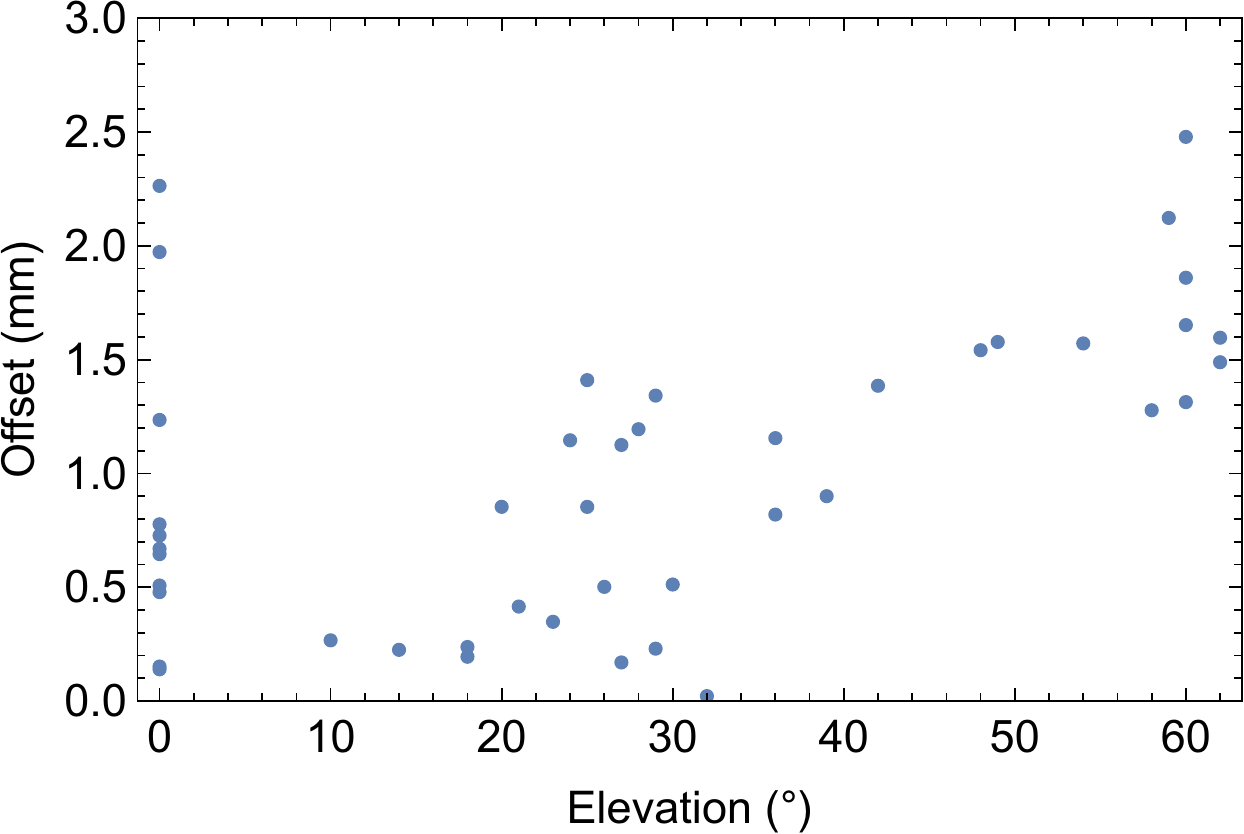}
  \caption{\emph{Left:} Deviation of the focal spot of the X-ray mirror during the X-Calibur flight as a consequence of mechanical and thermal bending of the truss. The red square indicates the nominal position and the blue dots show the positions measured during the flight. The diameter of the scintillator and a \SI{3}{\mm} radius around the center are indicated by the solid and dashed circles, respectively. \emph{Right:} Offset of the focal spot from the detector center as a function of pointing elevation. Measurements taken during ascent when the truss was latched are shown with an elevation of \SI{0}{\degree}. Thermal and mechanical deformations were smaller than \SI{1.5}{\mm} except during the ascent when the truss was latched and some periods of high-altitude pointing at \SI{60}{\degree} or more.}
  \label{fig:truss-deflections}
\end{figure}

\begin{figure}
  \centering
  \includegraphics[width=0.4\textwidth]{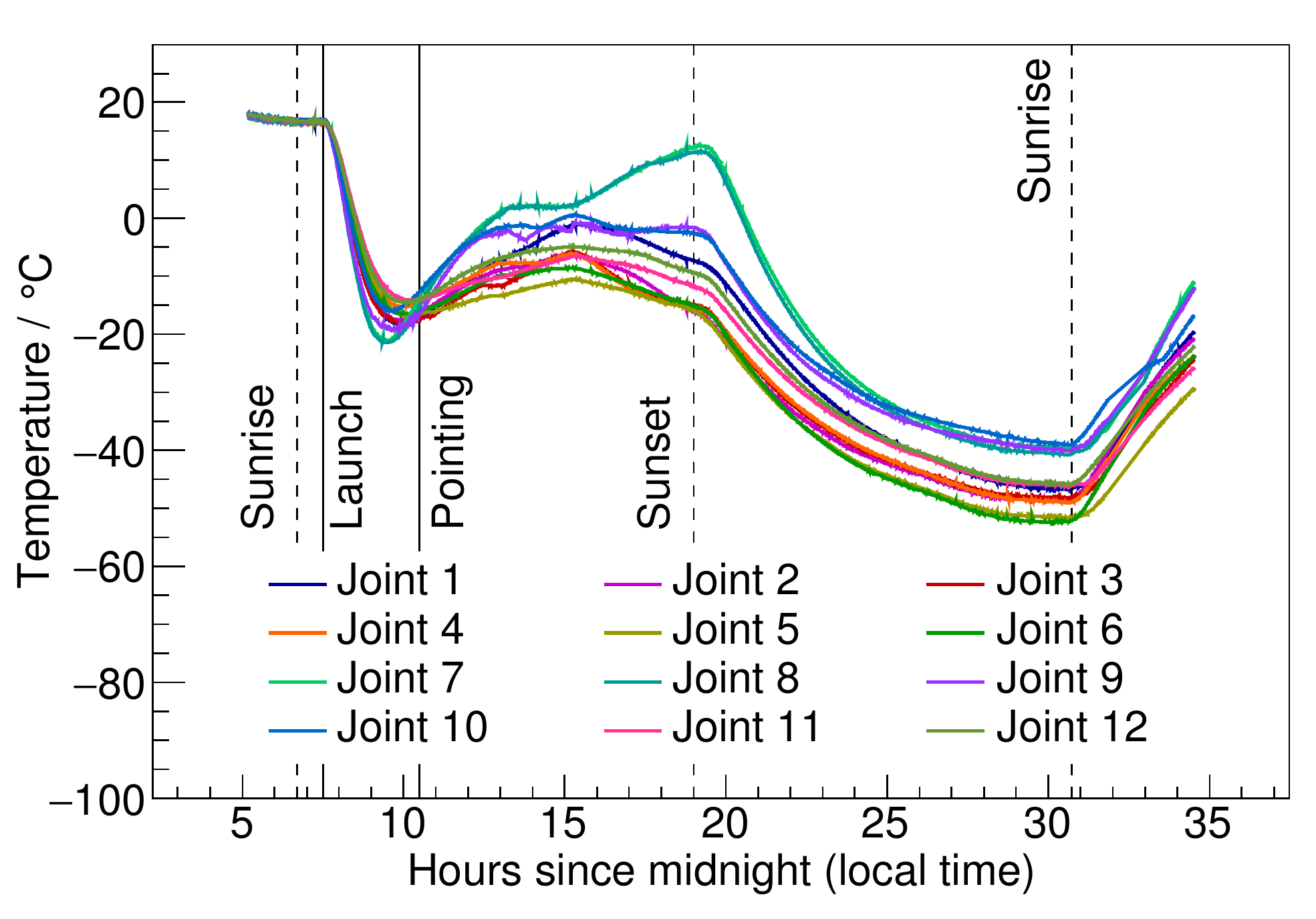}
  \caption{Temperatures measured on 12 joints throughout the flight.During most of the flight, the temperature gradient across the truss was less than \SI{15}{\celsius}. Joints 7 and 8 are noteworthy in that they show a temperature behavior that is significantly different from the other joints. This is explained by the fact that the latching mechanism is mounted to an aluminum bracket which mounts to these joints. This aluminum bracket closely couples the temperatures of the two joints. Additionally, it is not covered in Mylar, leaving the painted surface exposed. This resulted in the coldest temperatures on these joints during ascend, and the hottest temperatures during the day.}
  \label{fig:joint-temperatures}
\end{figure}

Truss deflections as determined from images acquired by the back-looking camera were smaller than \SI{1.5}{mm} except for a period during ascent when the telescope truss was latched and hence constrained and when pointing at very high elevations, as shown in Fig.~\ref{fig:truss-deflections}.
While deflections are slightly larger than the specification, they still result in a negligible ${<}1\%$ systematic error on the polarization fraction, when correcting for the (known) offset during the analysis of the data.
Because deflections were in this tolerable range, and because the temperature gradients on the truss were relatively small, the joint heaters were not used.
Joint temperatures throughout the flight are shown in Fig.~\ref{fig:joint-temperatures}.
Except for a few hours in the late afternoon, temperature gradients along the truss were less than \SI{15}{\celsius}.

It should be noted at this point, that alignment information is only available until 8:50pm local time.
The laser system failed during the launch.
The back-looking camera, which is an IP camera, could no longer be contacted by the flight computer after 8:50pm.
The latter failure is surprising because the temperature of the camera was in the range from \SI{10}{\celsius} to \SI{25}{\celsius} throughout the entire flight.
Both failures are still under investigation, and are going to be addressed prior to the next flight of X-Calibur.

\begin{wstable}
  \caption{Summary of the observation log of the 2016 X-Calibur flight.}
  \label{tab:observation-times}
  \begin{tabular}{@{}lll@{}}
    \toprule
    Source          & Observation Window & On-Source Time \\
                    & [Local Time]       & [Minutes]      \\
    \colrule
    Crab (day)      & 11:23 -- 12:21     & 58             \\
    Sco X-1 (day)   & 13:52 -- 15:07     & 75             \\
    Cyg X-1 (day)   & 15:20 -- 18:43     & 106            \\
    Cyg X-1 (night) & 23:19 -- 2:09      & 73             \\
    Crab (night)    & 2:36 -- 5:23       & 120            \\
    \botrule
  \end{tabular}
\end{wstable}

\begin{figure}
  \centering
  \includegraphics[width=0.4\textwidth]{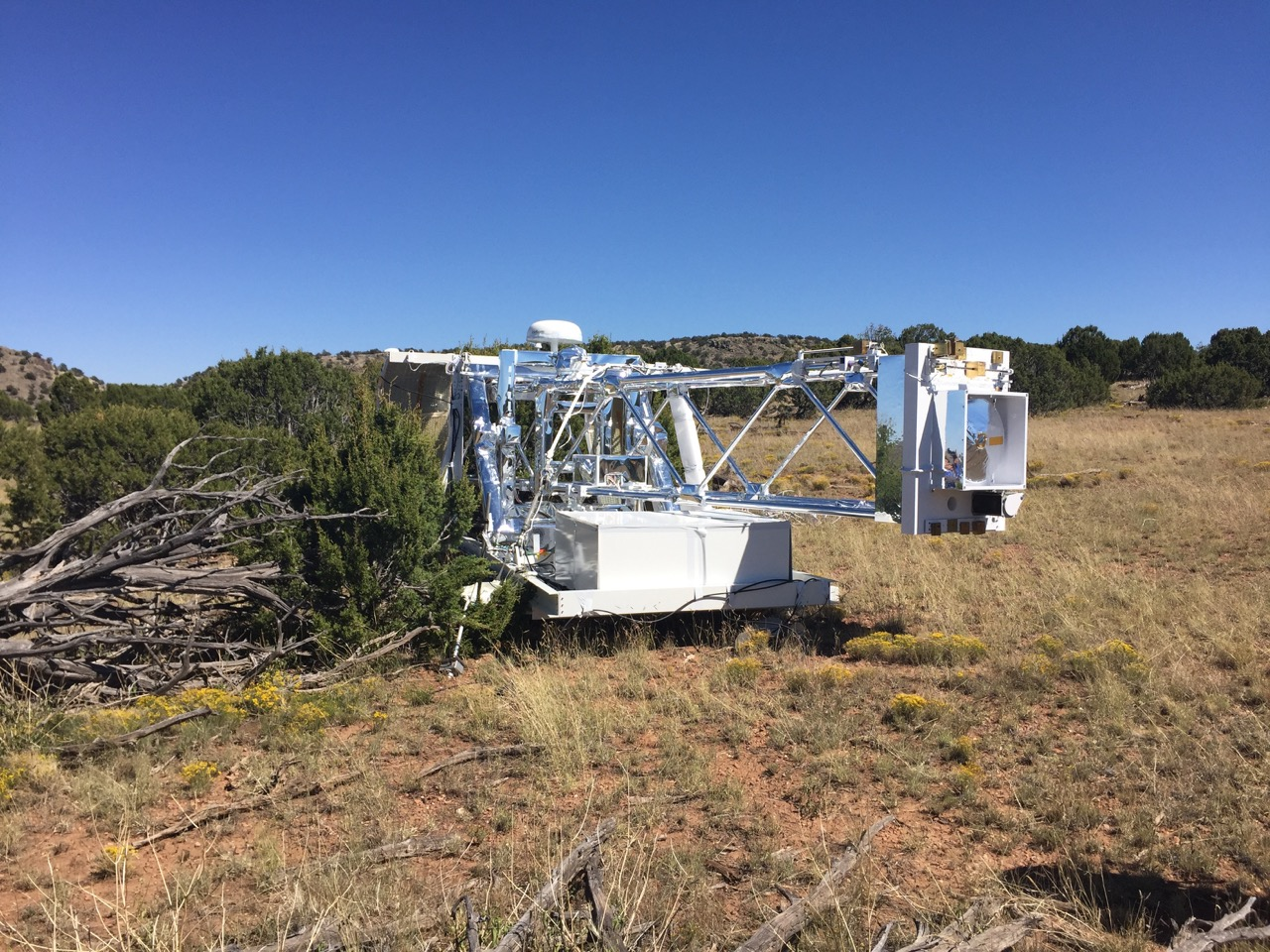}
  \caption{The X-Calibur balloon payload after landing.}
  \label{fig:recovery}
\end{figure}

During the flight, the Crab nebula, the accreting neutron star Sco X-1, and the accreting stellar mass black hole Cyg X-1 were observed.
On-source observation times are summarized in Table~\ref{tab:observation-times}.
The flight of X-Calibur was terminated 24.5 hours after reaching float altitude.
The X-ray mirror was recovered on September 18, only a few hours after landing.
The rest of the experiment was recovered on September 20.
Figure~\ref{fig:recovery} shows a photograph of the payload after landing.
During the landing, the telescope truss sustained only very minor damage on the surface of a few carbon fiber tubes.
After these damages have been repaired, the truss will be load-tested again, in order to verify that it still conforms with our stiffness requirements.
We expect to be able to reuse the truss for the next flight of X-Calibur from McMurdo, Antarctica, in 2018/19.
There was no damage to the polarimeter or any other components that were mounted on the truss.

\section{Summary and Outlook}\label{sec:summary}
Sensitive X-ray polarimetry has only become feasible relatively recently, with the development of affordable focusing X-ray optics.
X-Calibur is a balloon-borne hard X-ray scattering polarimeter, flown on a 1-day flight from Ft.\ Sumner, NM, on September 17th and 18th, 2016.
It consists of a telescope truss carrying an \SI{8}{m}-focal length X-ray mirror and the polarimeter in its focal plane.
During the balloon flight, the telescope is pointed at astrophysical objects with arcsecond precision by the Wallops Arc-Second Pointer (WASP).

One of the challenges during the development of the X-Calibur telescope truss was the fact that, in order to keep systematic uncertainties small, the focal spot of the mirror must be kept aligned with the scattering element of the polarimeter.
In this paper, we describe the design of the telescope truss, the tests we performed in the laboratory in order to ensure that it will meet our requirements, and the performance of the entire instrument during the flight.
We used two systems to measure the offset of the focal spot from the detector center during the flight.
Data from these systems showed that the truss met our stiffness requirements, achieving an alignment of \SI{1.5}{\mm} through most of the flight and an alignment knowledge better than \SI{1}{\mm}.
The WASP system pointed the telescope at three astrophysical sources (Crab nebula, Sco X-1, and Cyg X-1) with a tracking stability of about \SI{0.3}{\arcsecond} RMS.

The balloon gondola was recovered after the flight with only minor damages, which are currently being repaired.
There was no damage to the X-ray mirror, the polarimeter, and the components of the pointing system.
We plan to upgrade X-Calibur for a long-duration balloon flight from McMurdo, Antarctica, which is planned for the 2018/19 season.
Currently, the active CsI shield only covers the sides and the rear end of the polarimeter.
The front is only covered by a passive Tungsten shield.
An active CsI shield will be added at the front, which is expected to reduce backgrounds by a factor of 3.
Furthermore, an imaging CZT detector will be added behind the polarimeter.
About \SI{30}{\percent} of photons do not scatter in the scintillator.
The imaging detector will thus create a slightly out-of-focus image of the source.
This will allow us a more direct and immediate feedback of the mirror-to-detector and control-system-to-mirror alignments than the photons in the polarimeter or the other alignment monitoring systems.

\section*{Acknowledgements}
We are grateful for NASA funding from grants NNX10AJ56G, NNX12AD51G and NNX14AD19G, as well as discretionary funding from the McDonnell Center for the Space Sciences to build the X-Calibur polarimeter.
WASP is funded by the Balloon Program Office at NASA Wallops Flight Facility and NASA Headquarter.

\bibliographystyle{ws-jai}
\bibliography{truss}

\end{document}